\documentclass[11pt,letter]{article}
\usepackage[margin=1in]{geometry}

\usepackage{lipsum} 
\usepackage{graphicx}
\usepackage{amsmath}
\usepackage{caption}
\usepackage{subcaption}
\usepackage{lineno}
\usepackage{cuted}
\usepackage{hyphenat}
\usepackage{setspace}
\usepackage{color,soul}

\begin{document}

\title{\textbf{In vitro evaluation of solution-based pressurised metered dose inhaler sprays with low GWP propellants} }
\author{Daniel J Duke$^1$, 
Lingzhe Rao$^1$, 
Benjamin Myatt$^2$,\\
Phil Cocks$^2$,
Stephen Stein$^3$,
Nirmal Marasini$^4$,
Hui Xin Ong$^{4,5}$,
Paul Young$^{4,5}$
\vspace{5mm}\\
\small $^1$ Laboratory for Turbulence Research in Aerospace \& Combustion (LTRAC),\\
\small Department of Mechanical \& Aerospace Engineering, Monash University, Australia\\
\small $^2$ Kindeva Drug Delivery, Loughborough, United Kingdom\\
\small $^3$ Kindeva Drug Delivery, Woodbury, Minnesota USA\\
\small $^4$ Woolcock Institute of Medical Research, Glebe NSW 2037 Australia\\
\small $^5$ Department of Marketing, Macquarie Business School,\\ \small Macquarie University, Sydney, NSW 2109, Australia
}
\date{\small Preprint submitted to ArXiV - \today}
\maketitle

\section*{Abstract}

\textbf{Background:}
The impending transition from high Global Warming Potential (GWP) propellants such as HFA134a to low-GWP alternatives such as HFA152a and HFO1234ze(E) in pressurised metered dose inhalers (pMDIs) poses a number of challenges. Changes in chemicophysical prop- erties will alter spray performance; this study investigates those differences using matched hardware and formulations.
\\
\textbf{Methods and Materials:}
Aerodynamic particle size distribution (APSD) measurements, laser diffraction and high-speed imaging experiments were used to compare the performance of HFA134a, HFA152a and HFO1234ze(E) solution formulations of becamethasone dipropionate (BDP). Propellant-only placebos, cosolvent-free solutions (0.05 mg/mL), 8\% and 15\% w/w ethanol solution formulations  (2.0 mg/mL) were investigated.
\\
\textbf{Results:}
HFA152a formulations were found to have increased actuator and throat deposition while HFO1234ze(E) showed comparable APSD performance to HFA134a. Droplet and particle size increased for both low-GWP propellants. High-speed imaging revealed that HFA152a formulations spread more rapidly and were less stable and repeatable than HFA134a. HFO1234ze(E) formulations were found to spread more slowly than HFA134a, but converged with HFA134a ex-mouthpiece. Propellant effects were moderated by the addition of ethanol. 
\\
\textbf{Conclusions:}
Stability changes were found to be dominant in the near-orifice region, depending on formulation properties and ex-orifice flashing behaviour. Shot-to-shot repeatability effects were more pronounced in the ex-mouthpiece region where mixing with the ambient air is dominant. As such, modifications to orifice and mouthpiece geometry may provide a route to improved in vitro bioequivalence for low-GWP PMDIs.

\color{black}

\section{Introduction}

Choice of propellant plays a major role in the performance of solution-based pressurised metered dose inhalers (pMDIs) \cite{2002.Noakes,Smyth.2003}. Propellant properties dictate the structure of the plume, the size of the droplets formed and the properties of the particles produced through a range of complex chemical, thermodynamic and fluid-mechanical interactions \cite{Stein.2021,Myatt.2015,2006.Versteeg}. Due to the complexity of these processes, the mechanisms by which propellant properties affect the pharmaceutical performance of a product (i.e. fine particle fraction, droplet size, etc.) and thus drug delivery efficacy are not fully understood. \color{black}

The impending transition from hydrofluroalkane (HFA) propellants such as HFA134a and HFA227 to low global warming potential (low-GWP) propellants such as HFA152a and HFO1234ze(E) will involve significant changes in propellant chemico-physical properties such as vapour pressure and density\cite{DukeRDD2023}. Lack of understanding of the influence of these properties on pharmaceutical performance therefore poses a challenge for the design of effective solution-based pMDIs \cite{Pritchard.2020,2020.Pritchard}.\color{black}

The chemico-physical properties of the propellant not only influence the solubility and stability of the active pharmaceutical ingredient (API). They \color{black} also dictate the properties of the spray plume in the near-orifice region \cite{DukeRDD2023,2017.Mason-Smith}. 
The overall internal geometry of the actuator (such as the sump and valve stem volume, and ratio of metering valve to orifice diameter) act together with the formulation, particularly the propellant vapour and liquid densities, to determine the amount of internal vapourisation that initiates aerodynamic atomisation in the nozzle orifice \cite{Stein.2014}.
Propellant vapor pressure orifice size subsequently dictate \color{black} plume velocity and momentum which in turn controls the rate of droplet formation and mixing in the early regions of the spray (i.e. inside the mouthpiece) \cite{2017.Gavtash}. The surface tension of the formulation also drives initial droplet size \cite{Gavtash.2018}. The boiling point and  latent heat of vapourisation \color{black}of the propellant determine the degree to which the plume flash-evaporates when it exits the orifice; flashing is known to enhance atomisation and produce finer droplets \cite{2017.Gavtash}. All these factors influence the geometry of the spray plume and particle maturation \cite{Zhu.2013}, which influences throat deposition and fine particle fraction \cite{Stein.2021}.

\begin{table}[b!]
\centering
\begin{tabular}{|rccc|}   \hline 
\multicolumn{1}{|r|}{{  }}                 & \multicolumn{1}{c|}{{ \textbf{HFA134a}}}       & \multicolumn{1}{c|}{{ \textbf{HFA152a}}}       & { \textbf{HFO1234ze(E)}} \\ \hline
\multicolumn{1}{|r|}{{ Global warming potential }}             & \multicolumn{1}{c|}{{ 1430 kg CO$_2$e }}    & \multicolumn{1}{c|}{{ 124 kg CO$_2$e }}     & { 7 kg CO$_2$e }   \\ \hline 
\multicolumn{1}{|r|}{{ Saturation pressure at 25$^\circ$C}} & \multicolumn{1}{c|}{{ 6.7 atm}}       & \multicolumn{1}{c|}{{ 5.6 atm}}       & { 4.6 atm}      \\ \hline
\multicolumn{1}{|r|}{{ Liquid density}}             & \multicolumn{1}{c|}{{ 1207 kg/m$^3$}}    & \multicolumn{1}{c|}{{ 904 kg/m$^3$}}     & { 1170 kg/m$^3$}   \\ \hline
\multicolumn{1}{|r|}{{  Vapor density}}             & \multicolumn{1}{c|}{{  5.24 kg/m$^3$}}    & \multicolumn{1}{c|}{{  3.37 kg/m$^3$}}     & {  5.69 kg/m$^3$}   \\ \hline
\multicolumn{1}{|r|}{{ Jakob Number (eqn. \ref{eq:Ja})\color{black} }}             & \multicolumn{1}{c|}{{ 0.41 }}    & \multicolumn{1}{c|}{{ 0.32 }}     & { 0.36 }   \\
\multicolumn{1}{|r|}{{  (isenthalpic flashing potential)  }} & \multicolumn{1}{c|}{}  & \multicolumn{1}{c|}{\textit{(-23\%)}} & \multicolumn{1}{c|}{\textit{(-12\%)}} \\ \hline
\multicolumn{1}{|r|}{{ Orifice Reynolds Number $\mathrm{Re}_D$}}             & \multicolumn{1}{c|}{{ $3.5 \times 10^4$ }}    & \multicolumn{1}{c|}{{ $3.2 \times 10^4$ }}     & { $2.4 \times 10^4$ }   \\  
\multicolumn{1}{|r|}{{   (turbulence level, eqn. \ref{eq:Re}\color{black})  }} & \multicolumn{1}{c|}{}  & \multicolumn{1}{c|}{\textit{(-7\%)}} & \multicolumn{1}{c|}{\textit{(-30\%)}} \\ \hline
\multicolumn{1}{|r|}{{ Orifice Weber Number  $\mathrm{We}_D$  }}             & \multicolumn{1}{c|}{{ $13 \times 10^3$ }}    & \multicolumn{1}{c|}{{ $9.1 \times 10^3$ }}     & { $9.0 \times 10^3$ }   \\  
\multicolumn{1}{|r|}{{  (atomisation potential, eqn. \ref{eq:We}\color{black}) }} & \multicolumn{1}{c|}{}  & \multicolumn{1}{c|}{\textit{(-31\%)}} & \multicolumn{1}{c|}{\textit{(-32\%)}} \\ \hline
\multicolumn{1}{|r|}{{ Theoretical stable droplet size  }}             & \multicolumn{1}{c|}{{ 5.7 \textmu m }}    & \multicolumn{1}{c|}{{ 9.7 \textmu m \textit{(+70\%)} }}     & { 7.5 \textmu m \textit{(+32\%)} }  \\
\multicolumn{1}{|r|}{{ ( $d^*$, eqn. \ref{eq:WeMin})\color{black} }} &  \multicolumn{1}{c|}{} &  \multicolumn{1}{c|}{} &  \multicolumn{1}{c|}{}  \\ \hline 
\end{tabular}
\caption{Relevant propellant properties and experimental conditions. Changes for low-GWP propellants are given as percentages relative to HFA134a.\label{tab1}}
\end{table}

Unlike the prior CFC to HFA transition \cite{2016.Noakes}, the chemico-physical properties of the new low-GWP propellants are not as favourable with respect to vapour pressure, latent heat and density \cite{DukeRDD2023}. Some key properties are highlighted in Table \ref{tab1} (data from \cite{Katsuyuki.2016,1987.Sato,2007.Lemmon}). Both HFA152a and HFO1234ze(E) have  a lower propensity to flash-evaporate, as quantified by reduced Jakob number \cite{1998.Zeigerson-Katz}, the ratio of sensible to latent heat: \color{black}
\begin{equation}
\mathrm{Ja} = \frac{c_p \left( T_0 - T_{\mathrm{sat}} \right)}{\Delta h_{fg}}\label{eq:Ja} 
\end{equation}
While HFA152a has a higher vapor pressure  than HFA134a\color{black}, its density is much lower \cite{1987.Sato}. This yields low spray momentum, which will likely cause more rapid mixing. HFO1234ze(E) has a similar density similar to HFA134a but a lower vapor pressure \cite{Katsuyuki.2016}, thus reducing spray velocity, since for an ideal nozzle the velocity scales with pressure and density according to the following: \color{black}
\begin{equation}
\overline{U} \propto \sqrt { \frac{2 \left(P_{\mathrm{sat}} - P_{\mathrm{amb}} \right)}{\rho_{\mathrm{mix}}}}.\label{eq2}
\end{equation}
For both HFA152a and HFO1234ze(E) we therefore expect initial droplet sizes to increase. For HFA152a this is a density-driven effect and for HFO-1234ze(E) it is a velocity-driven (i.e. vapor pressure) effect.  The presence of cosolvent will likely moderate the differences between propellants, as has been previously observed in HFA134a and HFA152a systems \cite{Stein.2021}.

The miscibility of the propellant with other components of the formulation and the thermal and mass diffusivity properties of this complex mixture also likely play a role in drug particle formation and morphology \cite{1999.Vervaet,2019.Sheth}. It is well known that water content is a critical factor in particle formation \cite{2017.Ivey}, thus the solubility of water in the propellant and the ability of the spray plume to entrain humid surrounding air into the spray core has significant influence on particle maturation, residual droplet size and therefore region/location of deposition \cite{LeghLand.2021}. \color{black} Differences in turbulent mixing and entrainment between propellants can vary significantly from one shot to the next and occur at very short time and length scales due to high Reynolds and Weber number, as per Equations \ref{eq:Re}-\ref{eq:We}. For these reasons, they are not easily captured by conventional spray pattern and plume geometry methods \cite{Baxter.2022}.Typical values of Re and We relevant to the propellants considered in this study can be found in Table \ref{tab1}.\color{black}
\begin{eqnarray}
\mathrm{Re}_D =  \frac{ \rho \overline{U} D }{ \mu }\label{eq:Re}\\
\mathrm{We}_D = \frac{ \rho \overline{U}^2 D } {\sigma}\label{eq:We}
\end{eqnarray}

At present there is insufficient data to confirm the hypotheses stated above or propose modifications which may address the challenges posed by the low-GWP propellant switch. In this paper we have used a combination of conventional pharmacopoeial measurement approaches, laser droplet sizing and a novel ultra high speed plume imaging system to investigate the role of low GWP propellants HFA-152a and HFO-1234ze(E) on solution-based pMDI formulations using  conventional pMDI hardware \color{black} across a range of co-solvent concentrations. We show how subtle changes in the structure of the plume in the near-orifice and ex-mouthpiece region are correlated with changes in droplet and particle outcomes.

\section{Materials}

Twelve formulations were considered in this study, consisting of four API/cosolvent combinations in three propellants. These are summarised in Table \ref{tab2}. A standard set of pMDI hardware was used consistently across all formulations. The first set of formulations are propellant-only placebos of three propellants; HFA134a, HFA152a or HFO1234ze(E). The second set are low-dose solutions of 0.05 mg/mL  beclomethasone dipropionate (BDP) dissolved or solubilised in propellant only \color{black} as per Stein et al \cite{Stein.2021}. These were use to model aerodynamic particle size distribution outcomes in the limit of no cosolvent where propellant differences are expected to be largest. The third and fourth sets are conventional solution formulations of 2 mg/mL BDP in ethanol at 8\% and 15\% w/w respectively, in each of the three propellants. This large set of formulations permits consideration of the effects of low-GWP propellants on cosolvent-free formulations and observation of the moderating effect of cosolvent addition up to a typical concentration range. BDP was supplied by (Kindeva Drug Delivery, Loughborough, United Kingdom). Methanol, ethanol (100\%) and \color{black}other high-performance liquid chromatography (HPLC) grade solvents were supplied by Sigma-Aldrich (Castle Hill, New South Wales, Australia). Industrial-grade propellants (99.5\% purity) were supplied by A-Gas Australia.

A standard Kindeva pMDI actuator with 0.3 mm orifice diameter, 0.8mm orifice length and 14\,mm$^3$ sump volume was used for all the droplet sizing and imaging experiments. Formulations were prepared by weighing pre-micronised BDP using a precision balance (AS 62.R2+, Radwag) and adding ethanol to form a solution concentrate which was homogenised and dispensed into 16 mL FEP-coated aluminium canisters (Kindeva) prior to crimping. A 50 \textmu L metering valve (BK357, Bespak, UK) was crimped using a  Pamasol laboratory plant (2002, Pamasol, Switzerland) \color{black}and the canisters were filled with propellant to 200 nominal actuations. The propellant placebos, 8\% and 15\% ethanol containing \color{black}solutions were prepared at Monash University.  The cosolvent-free BDP solutions were prepared at Kindeva's facility in Loughborough (UK) following the same procedure however an Kindeva actuator with 0.65 mm orifice length was used for testing \cite{Stein.2021}.\color{black}

\begin{table}[]
\centering
\begin{tabular}{|rcccc|}
\hline
\multicolumn{5}{|c|}{{ \textbf{pMDI Formulations}}}                                                                                                                                                                \\ \hline
\multicolumn{1}{|r|}{{ Formulation }}     & \multicolumn{1}{c|}{ Propellant-only}    &  \multicolumn{1}{c|}{  Cosolvent-free} & \multicolumn{2}{c|}{ Ethanol-containing }       \\ 
\multicolumn{1}{|r|}{{}}       & \multicolumn{1}{c|}{  Placebo}    &  \multicolumn{1}{c|}{solution} & \multicolumn{2}{c|}{   solution formulations }       \\ \hline
\multicolumn{1}{|r|}{{ API}}                  &  \multicolumn{1}{c|}{  None }    & \multicolumn{3}{c|}{{ Beclomethasone Dipropionate (BDP)}}                   \\ \hline
\multicolumn{1}{|r|}{{ API concentration}}    &  \multicolumn{1}{c|}{ 0 mg/mL }        &  \multicolumn{1}{c|}{  0.05 mg/mL } &  \multicolumn{2}{c|}{ 2.0 mg/mL  }\\ \hline
\multicolumn{1}{|r|}{{ Co-solvent}}           & \multicolumn{1}{c|}{  None }      &  \multicolumn{1}{c|}{  None }    & \multicolumn{1}{c|}{  8\% w/w EtOH  }     & \multicolumn{1}{c|}{  15\% w/w EtOH  }           \\ \hline
\multicolumn{5}{|c|}{{ \textbf{Operating conditions}}}                                                                                                                                                            \\ \hline
\multicolumn{1}{|r|}{{ Ambient temperature}}        &  \multicolumn{4}{c|}{{ 21°C }}                                                                                                           \\ \hline
\multicolumn{1}{|r|}{{ Ambient humidity}}     &\multicolumn{4}{c|}{{ 50\% }}                                                                                                       \\ \hline
\multicolumn{1}{|r|}{{  Air flow rate}}         & 28.3 L/min    &  30.0 L/min & 28.3 L/min  & 28.3 L/min                                                                 \\ \hline
\multicolumn{1}{|r|}{{  Sample size (APSD)}}         & -    & $ n=24$ & $n=5$  & $n=5$                             \\ 
 \multicolumn{1}{|r|}{}     &     & $ N=3$ & $N=3$  &$N=3$                                                                 \\  \hline
\end{tabular}
\caption{Formulation properties and experimental conditions.\label{tab2}}
\end{table}

\section{Methods}

\subsection{Cascade Impaction APSD Measurements}

For the 2 mg/mL BDP 8\% and 15\% w/w ethanol solution formulations, an Andersen cascade impactor (ACI) was used to assess the aerosol performance and pulmonary deposition profiles of the pMDI formulations as per United States Pharmacopoeia guidelines (\S601). The measurements were performed at the Woolcock Institute (Sydney, Australia). For the 0.05 mg/mL cosolvent-free BDP formulations, testing was conducted at Kindeva Drug Delivery (Loughborough, UK) using a Next Generation Impactor (NGI, MSP Corporation, USA) operating at 30 L/min via a USP induction port with $n=6$ shots per configuration. 

ACI plates and NGI cups were coated prior to testing to ensure particle bounce was minimised. Multiple actuations were fired into the cascade impactor apparatus with a delay time between actuations to allow the pMDI to equilibriate. Data were processed using Copley Inhaler Testing Data Analysis Software (CITDAS, Copley Scientific, UK). Additional details are given in Table \ref{tab2} and further detail may be found in Duke et al \cite{DukeRDD2023}. 

\color{black}

The differences among groups were determined using one-way or two-way analysis of variance (ANOVA) followed by posthoc Tukey tests using GraphPad Prism 9. The difference was considered significant if the p-value was $\leq 0.05$.

\subsection{Laser diffraction droplet size measurement}

A volume-based particle size distribution was measured by laser diffraction using a Spraytec apparatus (Malvern Instruments, Worcestershire, UK) \cite{2004.Haynes}. Since the role of propellant on droplet size distribution is likely to be more pronounced closer to the actuator orifice, the measurements were taken immediately post-mouthpiece with the laser beam aligned to \color{black}the centre of the mouthpiece. Before testing, five \color{black}shots from each canister were fired to waste to ensure adequate valve priming. The particles were measured with a 300mm lens using a dispersion refractive index 1, particle refractive index 1.56, estimated density of 1.3 g/cm$^3$ and data acquisition rate of 10 kHz. Data from \color{black}$n=25$ repeated measurements were obtained for each formulation. Before testing, 5 shots from each canister were fired to waste to ensure adequate valve priming. Droplet size statistics were calculated during the steady period of the spray as defined by \color{black}conditionally averaging using the half-maximum of the peak laser absorption, with a cut-off filter at 46 \textmu m. \color{black}

\subsection{High speed plume imaging}
\label{method:imaging}

\begin{figure}
\centering
\includegraphics[width=.9\textwidth]{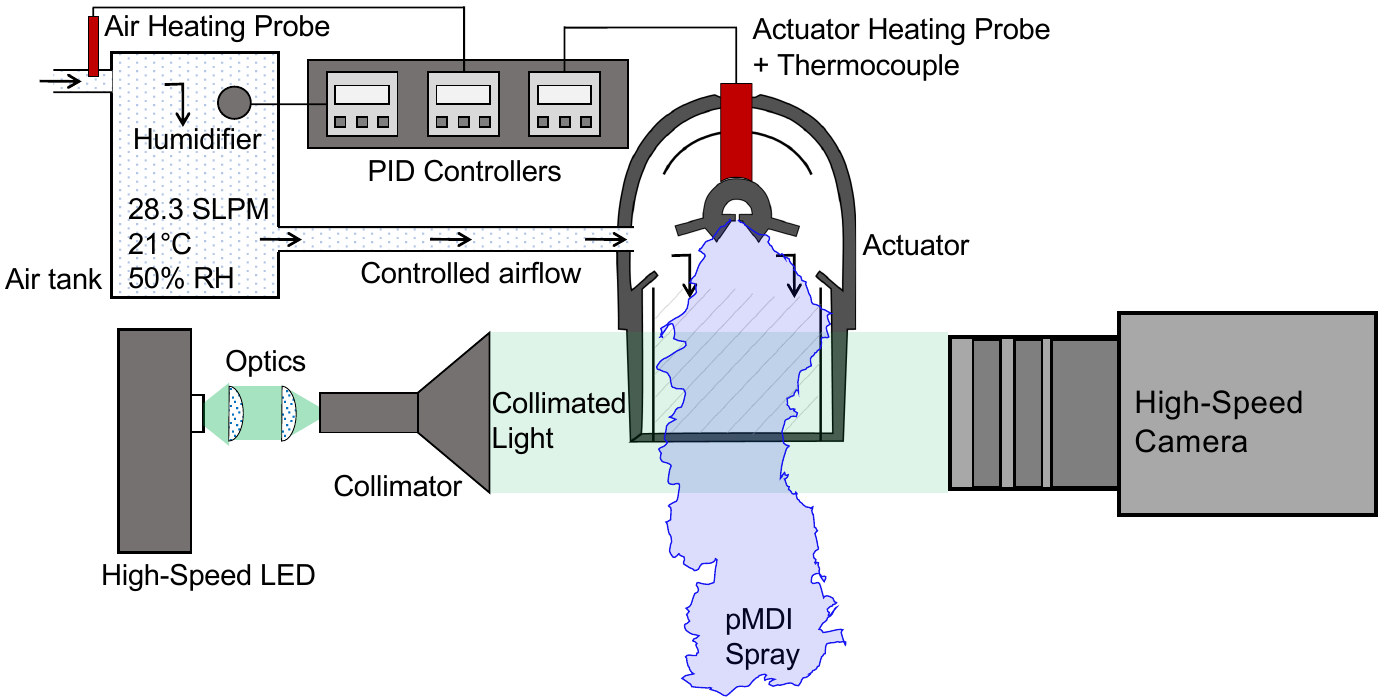}
\caption{Simplified schematic of the high-speed imaging system, viewed from above.\label{fig1}}
\end{figure}

In order to capture rapid temporal variations at small scales in the spray plume, a custom high-speed imaging system using diffuse backlit illumination was developed at Monash University \cite{DukeRDD2023}. A simplified diagram of the test facility is shown in Figure \ref{fig1}. A high speed camera (SA-Z, Photron, USA), 150mm macro lens (Nikon, Japan) and custom pulsed LED system \cite{2012.Buchmannr5r} (Monash University) with beam expander/collimator (Edmund Optics, Singapore) was placed either side of the mouthpiece to capture the structure of the plume with an effective exposure time of 350 ns at speeds of 20,000 to 140,000 frames/s with a fixed magnification of 31 \textmu m/pixel. 

Owing to the need to have an unobstructed view of the mouthpiece region, inhalation was simulated using an air conditioning system which delivered a constant flow of 28.3 standard L/min at 50\% relative humidity and \color{black}21$^\circ$C with positive pressure applied to the top of the actuator via a 3D printed adaptor. The adaptor was also fitted with a pneumatically operated plunger which depressed the aerosol \color{black}to trigger the spray with repeatable timing in synchronisation with the image capture\color{black}.  In order to prevent cooling and ice formation around the  actuator due to evaporative self-cooling of the formulation with repeated, \color{black}frequent actuations, a 15 W heating element was installed behind the nozzle block. The heater was used with a thermocouple and PID temperature controller to obtain constant nozzle block temperature throughout all tests. The timing between spray events was fixed at 25 s in order to minimise thermal effects. 

The presence of the mouthpiece and orifice exit cone impedes optical access to the orifice exit, where most primary droplet formation occurs \cite{2006.Versteeg}. However, removal of the mouthpiece can alter the plume structure as the mouthpiece constrains the periphery of the plume. To overcome this impediment, all experiments were repeated with three imaging configurations:
\begin{itemize}
\item Ex-orifice condition: \color{black}Mouthpiece removed and nozzle block machined back to partially remove the exit cone, permitting direct observation of the orifice exit plane ($N=60$ repeats of $n\approx 2000$ time-resolved snapshots per formulation).
\item Near-orifice condition: \color{black}Mouthpiece removed but nozzle block unmodified ($N=20$ repeats of $n\approx 6000$ time-resolved snapshots per formulation).
\item Ex-mouthpiece condition: Front and back sides of the mouthpiece removed, but the top and bottom left intact to preserve vertical constraint on the plume development ($N=15$ repeats of $n\approx 14000$ time-resolved snapshots per formulation).
\end{itemize}
In total, 10 TB of imaging data was obtained during the study. The data were analysed using MATLAB (Mathworks) software with in-house image processing algorithms.

\section{Results and Discussion}

\subsection{Impactor measurements}

\begin{figure}
\centering
\begin{subfigure}[b]{.8\textwidth} \centering
	\includegraphics[width=\textwidth]{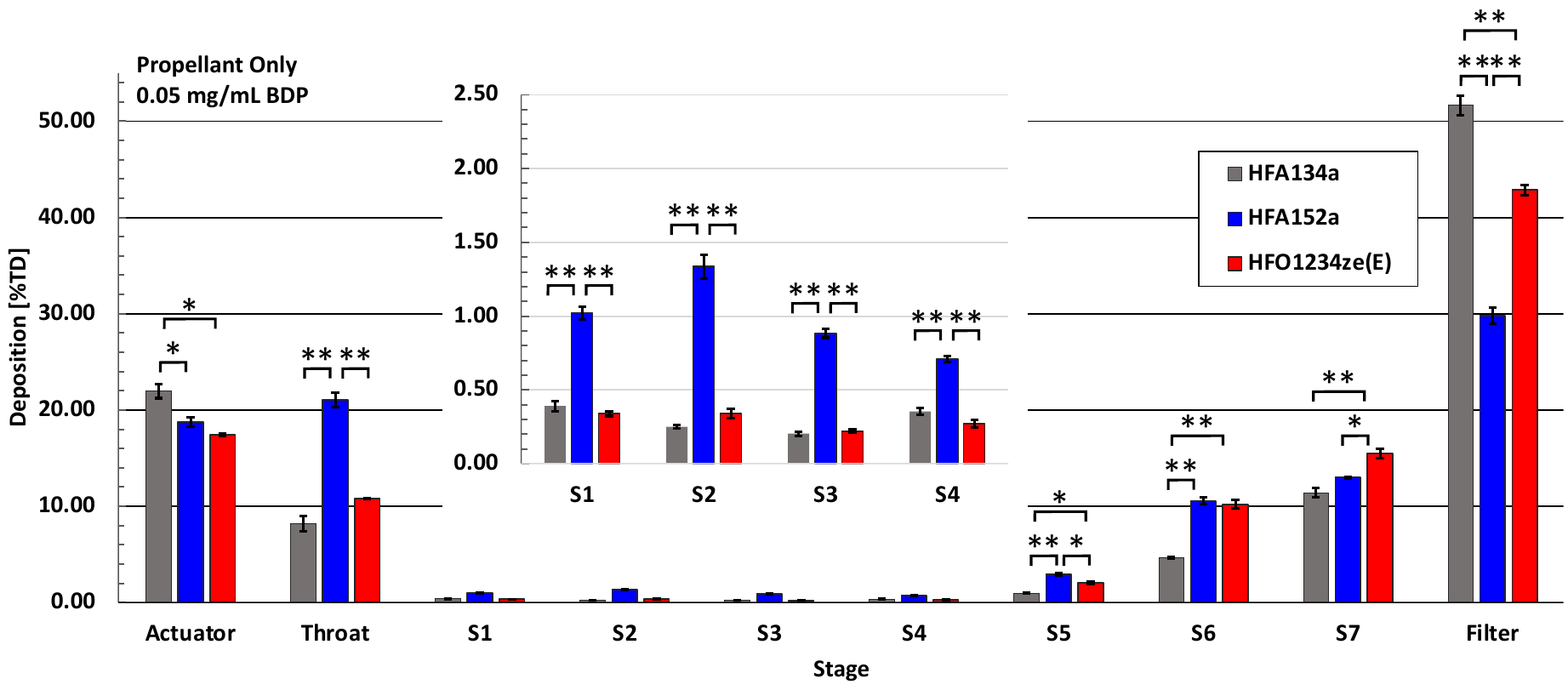}
	\caption{\label{fig2a}Next-generation impactor (NGI) data for propellant only \color{black}BDP formulations (0.05 mg/mL).}
\end{subfigure}
\begin{subfigure}[b]{.8\textwidth} \centering
	\includegraphics[width=\textwidth]{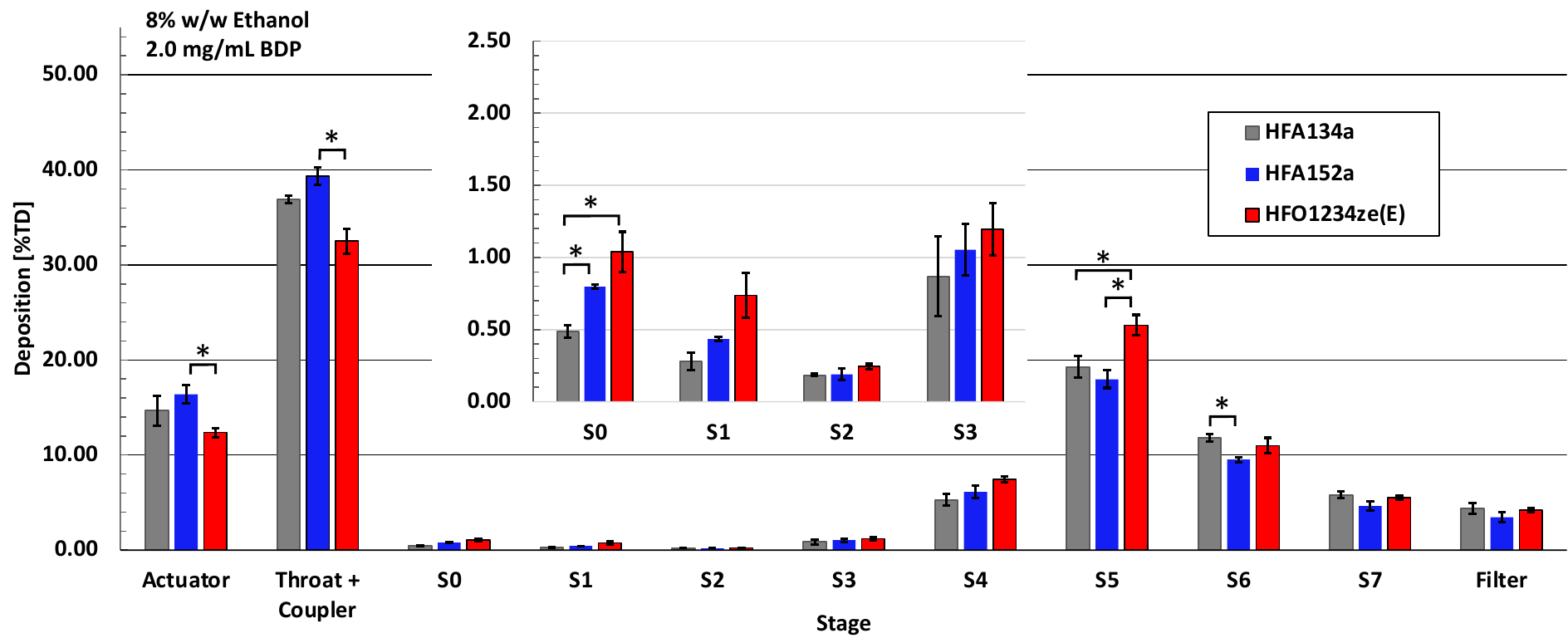}
	\caption{\label{fig2b}Andersen cascade impactor (ACI) data for 2.0 mg/mL BDP solution formulations (2.0 mg/mL) with 8\% w/w ethanol cosolvent.}
\end{subfigure}
\begin{subfigure}[b]{.8\textwidth} \centering
	\includegraphics[width=\textwidth]{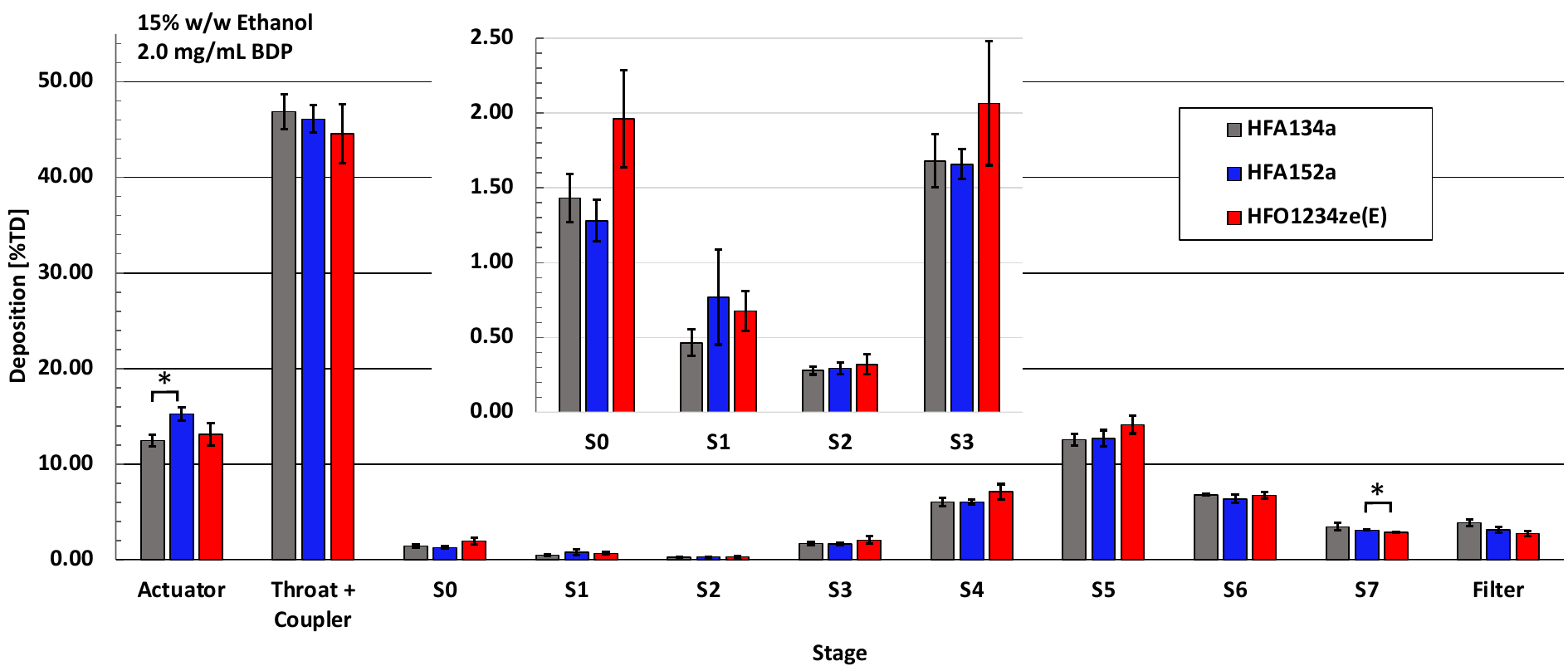}
	\caption{\label{fig2c}Andersen cascade impactor (ACI) data for 2.0 mg/mL BDP solution formulations (2.0 mg/mL) with 15\% w/w ethanol cosolvent.}
\end{subfigure}

\caption{\label{fig2}Aerodynamic particle size distribution results for BDP formulations.  The vertical axis is given as a percentage of total dose, ex-valve. \color{black} Significance indicators: * $p < 0.05$, ** $p < 0.01$.}
\end{figure}

Data from ACI (2 mg/mL BDP at 8\% and 15\% ethanol) and NGI (0.05 mg/mL BDP, 0\% ethanol) are shown in Table \ref{table:impactor}. Corresponding aerodynamic particle size distributions are shown in Figure \ref{fig2} for 0.05 mg/mL BDP (Figure \ref{fig2a}), 2 mg/mL BDP with 8\% ethanol (Figure \ref{fig2b}) and 2 mg/mL BDP with 15\% ethanol (Figure \ref{fig2c}). As expected,    differences in pharmaceutical performance metrics  (e.g. fine particle fraction, fine particle mass \& actuator deposition) \color{black} are largest for the cosolvent free formulations (Fig. \ref{fig2a}).  The addition of ethanol moderates the differences in propellant properties  and reduces \color{black}the statistical significance of the differences between formulations. A direct comparison with the ethanol-containing 2.0 mg/mL BDP solution formulations and the cosolvent-free formulation is not possible due to the differences in test protocol used (NGI vs ACI). However, protocols and test conditions were held constant between the three propellants, permitting a relative assessment of their performance for each set of experiments.

\begin{table}[]
\centering
\begin{tabular}{|r|ccc|}
\hline
{\textbf{Propellant}}                       & {\textbf{HFA134a}}    & {\textbf{HFA152a}}    & {\textbf{HFO1234ze(E)}} \\ \hline
{\textbf{Formulation}}                      & \multicolumn{3}{c|}{{\textbf{0.05 mg/mL BDP in Propellant only}}}                                                     \\
{{Total Dose Per Shot  {[}\textmu g{]}}}    & {2.73 ± 0.044}         & {2.77 ± 0.069}          & {2.01 ± 0.048 **}        \\
{{Calc. Delivered Dose  {[}\textmu g{]}}}    & {2.13 ± 0.016}         & {2.25 ± 0.069}         & {1.66 ± 0.042 **}        \\
{{Fine Particle Dose  {[}\textmu g{]}}}     & {1.89 ± 0.016}         & {1.59 ± 0.06 *}        & {1.43 ± 0.037 **}        \\
{{Fine Particle Fraction {[}\%{]}}}  & {89 ± 1.2}             & {71 ± 0.9 **}          & {86 ± 0.064}             \\
{{MMAD, GSD}}                        & {N/A}                  & {N/A}                  & {N/A}                    \\
 Actuator + throat deposition    & {30.16 ± 1.51}         & {39.80 ± 1.11 **}      & {28.23 ± 0.16}           \\
{[}\% of total API recovered{]} & & & \\ \hline
{\textbf{Formulation}}                      & \multicolumn{3}{c|}{{\textbf{2.0 mg/mL BDP in 8\% w/w Ethanol}}}                                                      \\
{{Total Dose Per Shot  {[}\textmu g{]}}}    & {99.57 ± 5.09}        & {97.18 ± 2.33}        & { 100.80 ± 3.40}          \\
{{Calc. Delivered Dose  {[}\textmu g{]}}}    & {84.86 ± 1.70}         & {81.21 ±0.40 \color{black} *}        & { 88.34 ±3.43 \color{black}*}         \\
{{Fine Particle Dose  {[}\textmu g{]}}}     & {47.24 ± 0.50}          & {41.62 ± 2.13 *}         & {53.55 ± 1.38 **}           \\
{{Fine Particle Fraction {[}\%{]}}}  & {55.69±1.67}           & {51.26 ±2.83}          & {60.67 ±2.66}            \\
{{MMAD {[}\textmu m{]}}}                    & {1.16 ± 0.00}          & {1.26 ±0.05}           & {1.29 ±0.05}             \\
{{GSD {[}\textmu m{]}}}                     & {1.74 ± 0.10}          & {1.78 ±0.01}           & {1.75 ±0.02}             \\
Actuator+coupler+throat deposition & {51.7 ± 1.96}          & {64.7 ± 2.14*}         & {52.6 ± 1.81}            \\
{[}\% of total API recovered{]} & & & \\ \hline
{\textbf{Formulation}}                      & \multicolumn{3}{c|}{{\textbf{2.0 mg/mL BDP in 15\% w/w Ethanol}}}                                                     \\
{{Total Dose Per Shot  {[}\textmu g{]}}}    & {97.19 ± 2.60}        & {101.97 ± 2.78}        & {103.00 ± 5.87}          \\
{{Calc. Delivered Dose  {[}\textmu g{]}}}    & {85.07 ± 1.90}         & {86.42 ± 2.03}        & {89.43 ± 3.64*}         \\
{{Fine Particle Dose  {[}\textmu g{]}}}     & {33.49 ± 2.69}         & {33.70 ± 2.86}        & {36.94 ± 5.73}          \\
{{Fine Particle Fraction {[}\%{]}}}  & {39.34±2.49}           & {38.97 ±2.94}          & {41.19 ±4.94}            \\
{{MMAD {[}\textmu m{]}}}                    & {1.34 ± 0.05}          & {1.40 ±0.03}           & {1.49 ±0.10}             \\
{{GSD {[}\textmu m{]}}}                     & {2.03 ± 0.07}          & {2.00 ±0.02}           & {1.99 ±0.03}             \\
Actuator+coupler+throat deposition & {63.5 ± 2.57}          & {64.7 ± 2.31}          & {61.3 ± 4.71}            \\
{[}\% of total API recovered{]} & & & \\ \hline
\end{tabular}
\caption{\label{table:impactor}Aerodynamic particle size distribution results for all propellant and formulation combinations tested.  Significance indicators: * $p<0.05$, ** $p<0.01$.}
\end{table}

The performance of HFA152a in the cosolvent free formulation is a clear outlier, showing increased deposition in the upper NGI stages (Fig. \ref{fig2a}), although the absolute values for these configurations are small (Table \ref{table:impactor}). Total dose per shot is similar to HFA134a but absolute fine particle dose is reduced by $15.9\% \pm 0.6\%$ relative to HFA134a. Although the APSD result for HFO1234ze(E) is comparable to HFA134a, the total dose per shot is notably reduced at $25.9\% \pm 0.7\%$ relative to HFA134a, with absolute fine particle dose (FPD) reduced by $24.3\% \pm 0.7\%$ relative to HFA134a. These results indicate that in the absence of cosolvent, HFA152a performance is impeded by reduced fine particle fraction, whilst HFO1234ze(E) performance is impeded by poor delivery efficacy. 

APSD results for the ethanol-containing 2.0 mg/mL BDP solution formulations (Figs. \ref{fig2b}-\ref{fig2c}) are relatively similar. 
 The data in Table \ref{table:impactor} shows that the addition of ethanol to the formulation reduces the differences in performance between the different propellant systems. moreover, the performance differences between propellant systems reduced as ethanol concentration in the formulation was increased. \color{black}However, differences between formulations are still evident in the absolute values. At 8\% w/w ethanol, HFA152a BDP formulations show a reduction in fine particle dose of $-11.9\% \pm 0.6\%$ relative to HFA134a, while HFO1234ze(E) formulations show an increase of $+13.3\% \pm 0.3\%$  for a comparable formulation composition \color{black}. At 15\% w/w ethanol, the HFA152a  FPD measured was similar to that of HFA134a whereas the FPD for the HFO1234ze(E) formulation was $+10.3\% \pm 1.8\%$ higher. \color{black}

Another noteworthy point of comparison between propellants is the actuator, coupler and throat deposition, which is $25\% \pm 2\%$ higher for HFA152a than for HFA134a in both the cosolvent-free and 8\% w/w ethanol formulations. This effect reduces at the higher cosolvent fraction. 
No such deposition change is noted for HFO1234ze(E) under any condition. 

HFA152a and HFO1234z(E) formulations with ethanol showed increased mass median aerodynamic diameter (MMAD) \color{black} relative to HFA134a, on the order of 9-11\% ($\pm 1\%$) , as per Table \ref{table:impactor}. MMAD could not be determined for the cosolvent-free formulations.  Increased MMAD is likely tied to increases in initial \color{black}droplet size and altered maturation timescales.   The fundamental propellant properties of \color{black}both HFA152a and HFO1234ze(E) lead to larger initial droplet size and reduced mixing which support increased MMAD. 

\subsection{Droplet sizing}

\begin{figure}
\centering
\includegraphics[width=.9\textwidth]{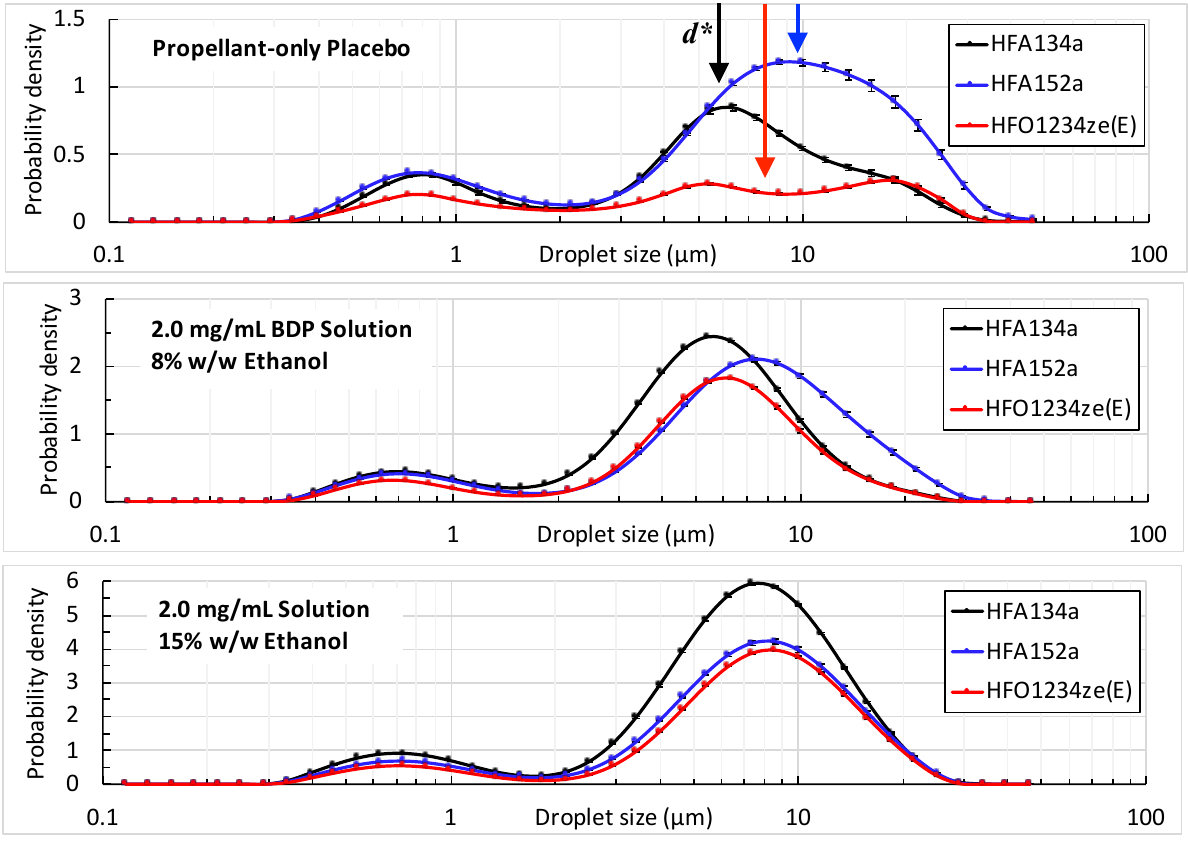}
\caption{\label{fig3}Laser droplet sizing data from Malvern Spraytec for propellant-only placebo formulations (top), 8\% and 15\% w/w BDP solutions (middle, bottom). The vertical arrows represent the theoretical stable droplet size $d^*$ (Equation. \ref{eq:WeMin}).}
\end{figure}

 In order to understand the causal factors that contribute to the trends observed in the APSD data with respect to particle size and delivered dose, measurements of droplet properties closer to the orifice are required. To this end, \color{black} ex-mouthpiece droplet size distributions were obtained  using laser diffraction \color{black}for the propellant-only placebo, and 2.0 mg/mL BDP solutions with 8\% and 15\% w/w ethanol. Time-average normalized size distributions during the steady period of the spray (conditional on laser absorption exceeding half the maximum value) are shown in Figure \ref{fig3}.  A quantitative comparison of the size distribution data is given in Table \ref{table:droplet}. Droplet size increases relative to the HFA134a control case \color{black}for both low-GWP propellants in all configurations, with the addition of ethanol in the formulation having a moderating effect. These changes can be mostly explained by changes in liquid density $\rho$ and surface tension $\sigma$ which define a minimum stable droplet size $d^*$ for a given spray. The value of $d^*$ can be estimated from the critical Weber number \cite{Hickey}:  \color{black}
\begin{equation}
\mathrm{We}^* = \frac{\rho \overline{U}^2  d^*}{\sigma} \approx 1\label{eq:WeMin}
\end{equation}
Indicative values of $d^*$ for propellant-only placebos are given in Table \ref{tab1}. 
 The theoretical and observed droplet sizes are best compared by considering the equivalent average diameter of spherical droplets having the same volume to surface area ratio ($V/S$) as that recorded in the spray with probability distribution $q(d)$ \cite{Kowalczuk}. This is described by the Sauter mean diameter ($d_{32}$) which is defined as:
\begin{equation}
d_{32} = \frac{6V}{S} = \frac{\int d^3 q(d) \, \mathrm{d}  d}{\int d^2 q(d) \, \mathrm{d} d} \label{eq:d32}
\end{equation}
\color{black}
The correlation between predicted droplet size from the chemicophysical properties only and the measured droplet size is $R^2 = 0.81$ for $d_{32}$ and $R^2 = 0.86$ for droplet MMAD. This indicates that most of the change in droplet size (Table \ref{table:droplet}) is driven by chemicophysical property changes rather than any change in the plume geometry or mixing.

\begin{table}[]
\centering
\begin{tabular}{|cccc|}
\hline
{ \textbf{Propellant}}                    & { \textbf{HFA-134a}} & { \textbf{HFA-152a}} & { \textbf{HFO-1234ze(E)}} \\ \hline
\multicolumn{4}{|c|}{{ \textbf{Pure-Placebo Formulation}}} \\
{ Sauter Mean Diameter $d_{32} $ {[}\textmu m{]}} & { 15.58 ± 0.56}      & { 19.45 ± 0.39}      & { 18.85 ± 0.22}           \\
{ Droplet MMAD {[}\textmu m{]}}                          & { 17.36 ± 0.63}      & { 20.90 ± 0.41}      & { 19.87 ± 0.18}           \\ \hline
\multicolumn{4}{|c|}{{ \textbf{2.0 mg/mL BDP in 8\% w/w Ethanol}}}\\
{ Sauter Mean Diameter $d_{32}$ {[}\textmu m{]}} & { 11.00 ± 0.09}      & { 14.76 ± 0.22}      & { 11.40 ± 0.11}           \\
{ Droplet MMAD {[}\textmu m{]}}                          & { 11.99 ± 0.12}      & { 16.04 ± 0.24}      & { 12.30 ± 0.14}           \\ \hline
\multicolumn{4}{|c|}{{ \textbf{2.0 mg/mL BDP in 15\% w/w Ethanol}}}\\
{ Sauter Mean Diameter $d_{32}$ {[}\textmu m{]}} & { 13.18 ± 0.11}      & { 13.82 ± 0.07}      & { 13.63 ± 0.04}           \\
{ Droplet MMAD {[}\textmu m{]}}                          & { 13.95 ± 0.12}      & { 14.69 ± 0.08}      & { 14.40 ± 0.04}          \\ \hline
\end{tabular}
\caption{\label{table:droplet}Ex-mouthpiece droplet size statistics  as determined by laser diffraction.}
\end{table}

\subsection{High speed plume imaging - time average statistics}

\begin{figure}
\centering
\includegraphics[width=\textwidth]{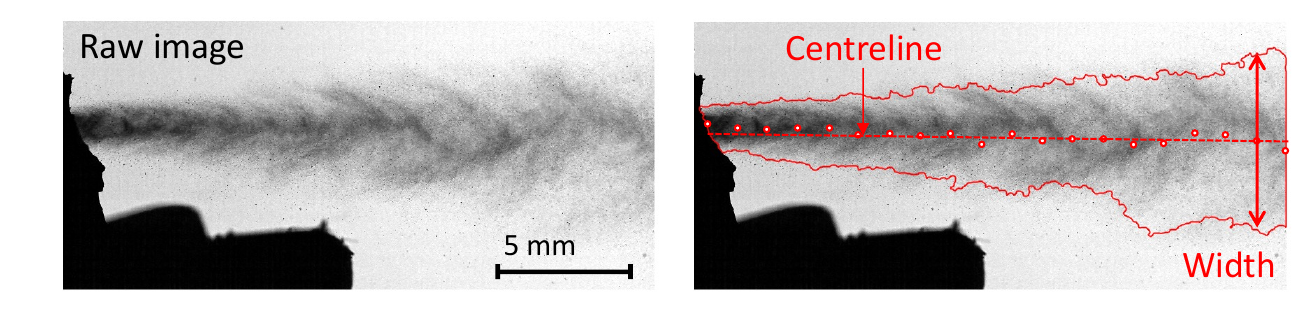}
\caption{\label{fig4}Sample plume image (left) and definition of spray width (right).}
\end{figure}

Changes in APSD with propellant are more complex than droplet size alone, and cannot easily be explained by simple changes in formulation properties. \color{black} Unsteady variations in plume geometry and mixing are expected to play a role. To investigate this, high speed imaging was used to measure the transient width, angle and intensity of the spray plume. An example high-speed image of a HFA134a plume ex-orifice with mouthpiece removed is shown in Figure \ref{fig4} (left). A custom MATLAB algorithm was developed to detect the spray boundary and centerline in each image at multiple streamwise locations, as per Figure \ref{fig4} (right). A spray width profile was developed for each shot as a function of both time and distance from the orifice.

The spray width data was ensemble-averaged over many repeated shots (  as per Section \ref{method:imaging} \color{black} ) and time-averaged over the steady period of the spray  to obtain a time-average, ensemble-average mean and standard deviation width profile for each formulation. The steady period of the spray was defined by the product of the coefficients of variance of the spray targeting angle and the volume-integrated light extinction in a region adjacent to the orifice. All imaging results are conditionally averaged on this product exceeding the half maximum value. \color{black} 

\begin{figure}
\centering
\begin{subfigure}[b]{\textwidth} \raggedright
	\includegraphics[height=5cm]{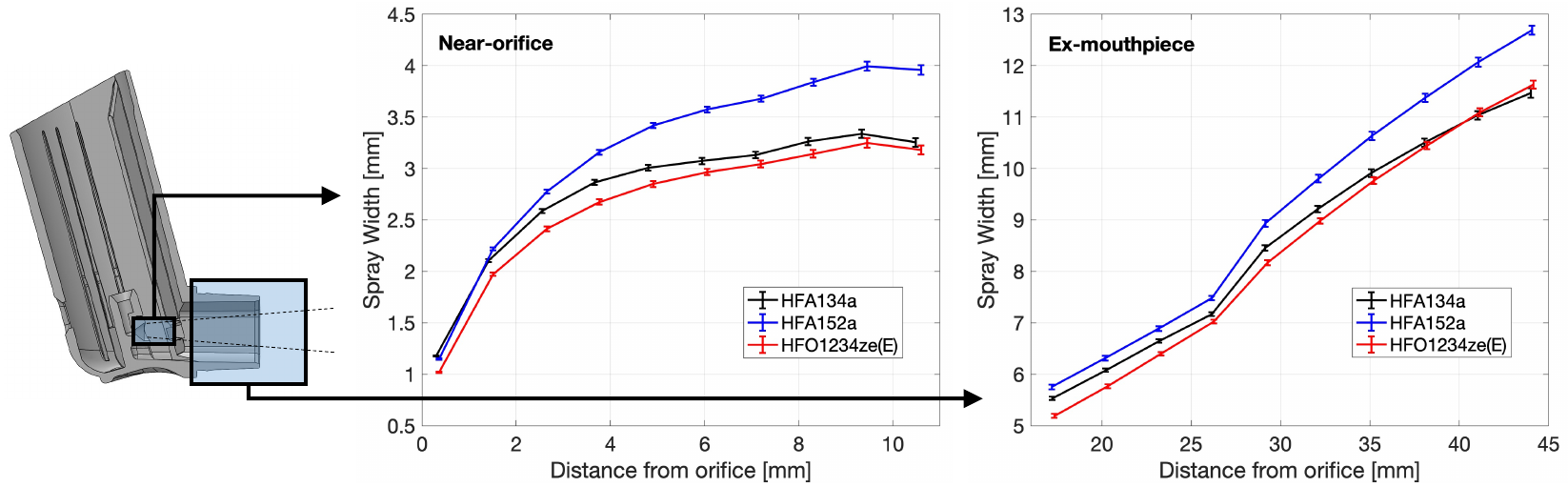}
	\caption{\label{fig5a}Spray width for propellant-only placebo formulations.}
\end{subfigure}
\begin{subfigure}[b]{\textwidth} \raggedleft
	\includegraphics[height=5cm]{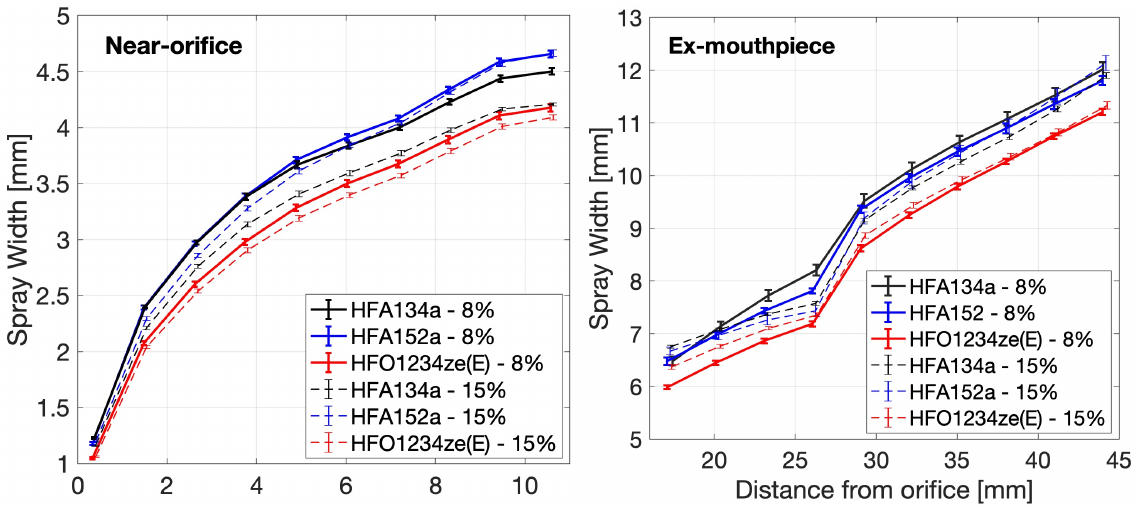} \hspace{10mm}
	\caption{\label{fig5b}Spray width for BDP solution formulations.}
\end{subfigure}
\caption{Spray width near-orifice (left) and ex-mouthpiece (right).\label{fig5} Error bars indicate  a combined standard deviation including both temporal and shot-to-shot variations.\color{black}}
\end{figure}

The spray width profiles are shown for all formulations in Figure \ref{fig5}. The measurements were taken both near-orifice (left) and ex-mouthpiece (right). The propellant-only placebos (Figure \ref{fig5a}) show HFA152a having a wider plume and HFO1234ze(E) a narrower plume relative to HFA134a. This is consistently observed from 1 mm (3  orifice \color{black} diameters) from the orifice, and extends throughout the domain. The shape of the average width profiles are comparable for all cases. As ethanol is added to the formulation (Fig. \ref{fig5b}) the  differences in plume width between propellant systems reduce but follow \color{black} the same trend.


\begin{figure}
\centering
\includegraphics[height=5cm]{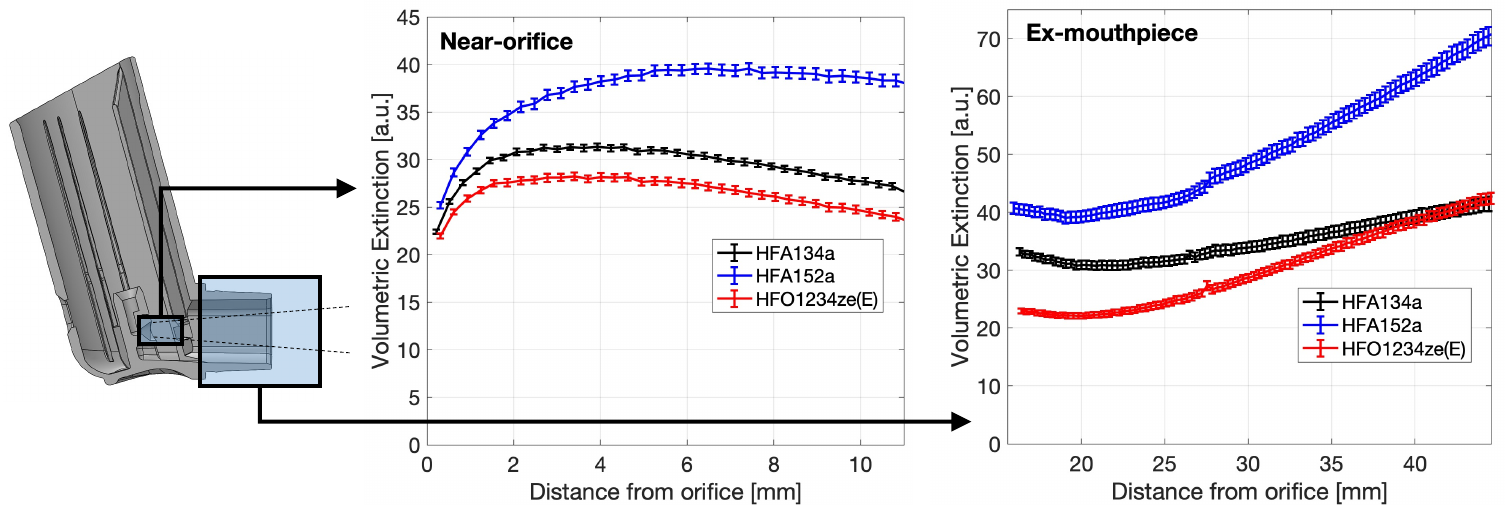}
\caption{\label{fig6} Volumetric extinction profiles $\left<\overline{I}_V\right>(x)$ for propellant-only placebos. Error bars indicate  a combined standard deviation including both temporal and shot-to-shot variation. \color{black} Results for BDP solutions available in supplementary material.}
\end{figure}

 In addition to plume width, light extinction in the spray provides insight into the number density of droplets and how this changes across formulations. The effect of changing droplet number density was determined by calculating a volume-averaged spray width profile. This was obtained by integrating the light extinction normalized by the incident intensity across the plume width. Plume width was \color{black} defined by the spray radius $r$ about its centerline $r_0$ during the steady period of the spray ($t_0$ to $t_1$):
\begin{equation}
\overline{I}_{V}(x) = \frac{1}{t_1-t_0} \int_{t_0}^{t_1} \int_{r_0(x)-r(x)}^{r_0(x)+r(x)} I \, \mathrm{d} r \, \mathrm{d} t \label{eq:vol}
\end{equation}
This calculation was ensemble-averaged over many repeated shots to obtain an ensemble-average, time-average volumetric extinction $\left<\overline{I}_V\right>(x)$ which is shown in Figure \ref{fig6}.  The overbar denotes time averaging, and the angle bracket denotes ensemble averaging (from shot to shot). HFA152a formulations have a consistently higher volumetric extinction relative to HFA134a, and HFO1234ze(E) has a consistently lower value relative to HFA134a. The spatial development of the volumetric extinction is similar across all propellants. The HFA152a profile diverges positively in the ex-mouthpiece region, indicating a wider spray relative to HFA134a. The HFO1234ze(E) profile remains below the HFA134a values in the near-orifice region and converges with the HFA134a profile ex-mouthpiece, indicating a narrower initial spray that is diluted by mixing with the ambient air further downstream.  


Plume behaviour in the near-orifice region dominated by initial flash-evaporation of the superheated propellant \cite{Gavtash.2016}, while behaviour further downstream is dominated by mixing behaviour with the ambient air \cite{2014.Buchmann}.  Expansion of vapour bubbles in the plume during the flashing process causes the spray width to increase rapidly near the orifice. As such, we expect the initial divergence of the spray width and volumetric density to depend inversely on the vapour phase density at atmospheric pressure (vapour bubbles of a given mass will expand to a greater size if density is lower). The spray width and volumetric extinction data presented in Figures \ref{fig5}-\ref{fig6} match this hypothesis; the initial divergence of between formulations (i.e. Figure \ref{fig5a}) correlates inversely with the vapour density (Table \ref{tab1}). HFA152 has a 36\% reduction in vapour density relative to HFA134a, and the plume is observed to expand to a correspondingly larger ratio. HFO1234ze(E) has a 9\% increase in vapour density relative to HFA134a, and its plume expands by a correspondingly lesser amount. This effect explains the near-orifice behaviour of HFA152a in particular.

Further downstream in the ex-mouthpiece region (Figure \ref{fig5b}) it is the average liquid-vapour mixture density rather than the vapour density alone that determines mixing. Here, vapour expansion is complete, the droplet field is dilute \cite{DukeUSAXS} and the plume behaves like a dense turbulent jet\cite{2014.Buchmann}. Assuming isenthalpic expansion at the orifice, the HFA152a has a 17\% reduction in mixture density relative to HFA134a. We therefore expect the plume to expand and decelerate by a corresponding amount and this agrees with the volumetric extinction profile in Figure \ref{fig6}. Conversely, HFO1234ze(E) has a 2\% increase in mixture density relative to HFA134a. Allowing for uncertainty due to thermal effects, this change is not statistically significant. As expected, we see HFO1234ze(E) and HFA34a converge to a similar volumetric extinction after the mouthpiece in Figure \ref{fig6}.
\color{black}

\subsection{High speed plume imaging - stability and repeatability}

So far, we have considered only the average behaviour of the plume. \color{black}To explore the differences between formulations in greater detail we consider the spatial structure of the plume and how this varies with both time and from shot to shot. Definitions of all mathematical symbols and variables may be found in the Appendix. \color{black}Firstly, the time-average  extinction $\overline{I}$ \color{black}during the steady period of the spray and the ensemble-average $\left< \overline{I} \right>$ \color{black}over $i=1$ to $N$ repeated sprays are defined as:
\begin{eqnarray}
\overline{I}_i = \frac{1}{t_1-t_0} \sum_{t=t_0}^{t=t_1} I_i (x,r,t)\\
\left< I \right> = \frac{1}{N}  \sum_{i=1}^{i=N} I_i (x,r,t).
\end{eqnarray}
These can be defined to create the time-average, ensemble-average mean extinction profile from which we have defined the spray width and volumetric extinction profiles above (Figures \ref{fig5}-\ref{fig6}):
\begin{equation}
\left< \overline{I} \right> = \frac{1}{N \left(t_1-t_0 \right)} \sum_{i=1}^{i=N} \sum_{t=t_0}^{t=t_1} I_i \label{eq:mean}
\end{equation}

We now define the temporal stability of the spray as the sample standard deviation in time ($s_i$) \color{black} of each individual spray during its steady period, which is then ensemble-averaged to obtain:
\begin{equation}
\left< \overline{I}_{stab} \right> = \frac{1}{N} \sum_{i=1}^{i=N} s_i = \frac{1}{N} \sum_{i=1}^{i=N} \sqrt{ \frac{1}{t_1-t_0}  \sum_{t=t_0}^{t=t_1} \left[ I_i - \overline{I}_i \right]^2} \label{eq:stab}
\end{equation}

Similarly, we can define the shot-to-shot repeatability of the spray as the sample standard deviation at each instant in time across $N$ repeated measurements ($s_t$) \color{black}, which can then be time-averaged to obtain:
\begin{equation}
\left< \overline{I}_{rep} \right> = \frac{1}{t_1-t_0} \sum_{t=t_0}^{t=t_1}  s_t = \frac{1}{t_1-t_0} \sum_{t=t_0}^{t=t_1} \sqrt{ \frac{1}{N-1} \sum_{i=1}^{i=N}\left[ I_i - \left< I \right> \right]^2} \label{eq:rep}
\end{equation}


\begin{figure}
\centering
\begin{subfigure}[b]{\textwidth} \centering
	\includegraphics[height=8cm,clip=true,trim=1cm 7mm 7mm 1cm]{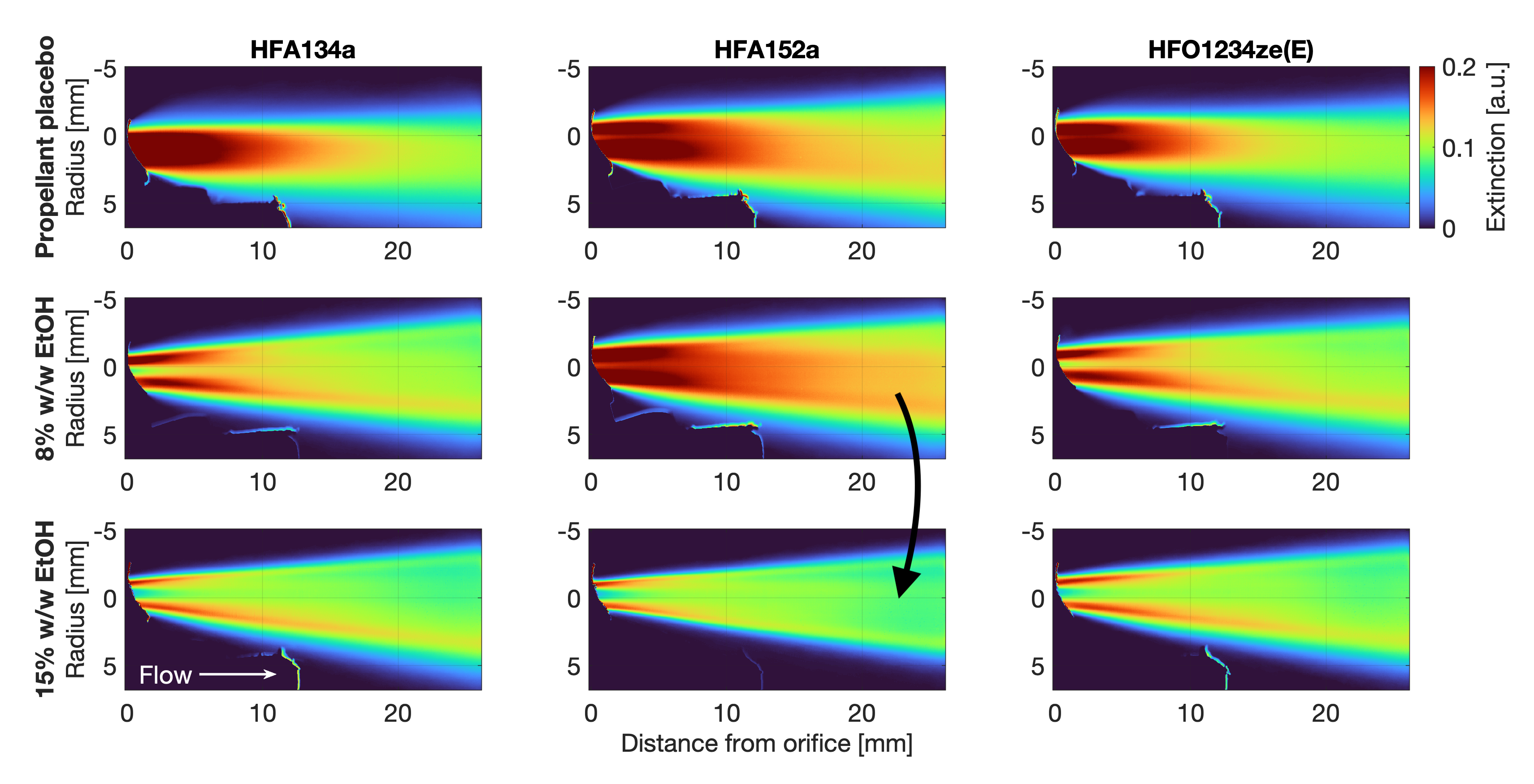}
	\caption{\label{fig7a}Spray temporal stability $\left< I_{stab} \right> (x,r)$}
\end{subfigure}
\begin{subfigure}[b]{\textwidth} \centering
	\includegraphics[height=8cm,clip=true,trim=1cm 7mm 7mm 1cm]{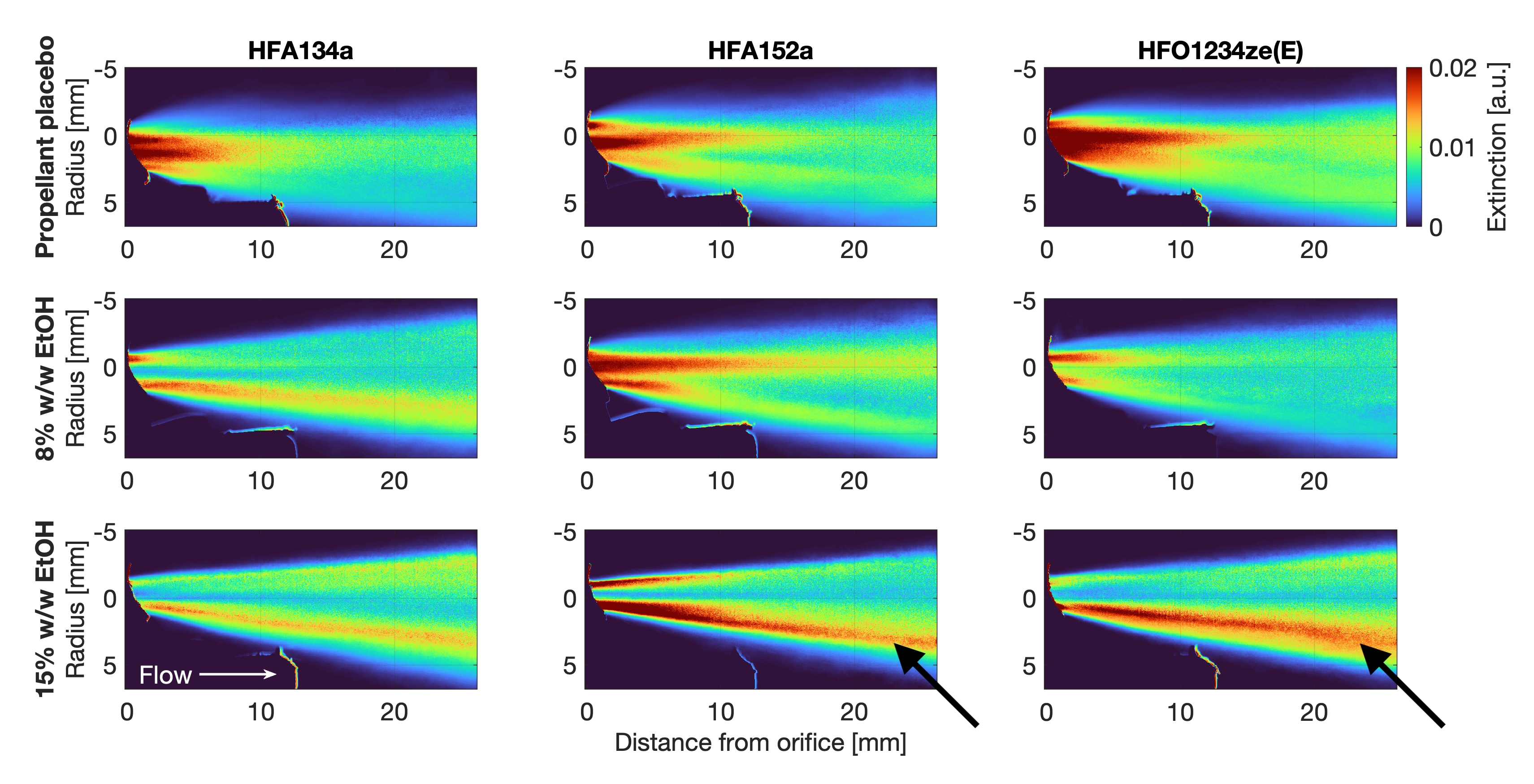}
	\caption{\label{fig7b}Shot-to-shot repeatability $\left< I_{rep} \right> (x,r)$}
\end{subfigure}
\caption{\label{fig7}Spatial structure of plumes for all formulations  in the near-orifice region \color{black}, showing (a) stability (eqn. \ref{eq:stab}) and (b) repeatability (eqn. \ref{eq:rep}). Flow is left-to-right. The horizontal axes are distance from orifice ($x$) and vertical axes are radius ($r$), in mm.}
\end{figure}

The mean intensity profiles (eqn. \ref{eq:mean}) are unremarkable, and do not exhibit any large differences between formulations beyond those shown in Figures \ref{fig5}-\ref{fig6}. These can be found in the supplementary material. The temporal stability profiles are shown in Figure \ref{fig7a}. Here, large values (red) indicate large temporal fluctuations and small values (blue) indicate no change over time. Adding ethanol has the effect of stabilising the spray in time, which is evident by the reduction in red regions in the core of the spray in the lower rows of images. However, far more ethanol is required to achieve the same effect for HFA152a than for the other propellants. At 8\% w/w ethanol both HFA134a or HFO1234ze(E) show significant reduction in temporal variability.  HFA152a requires the addition of 15\% ethanol to the formulation to achieve a reduction to a comparable level of temporal variability across the propellant formulations. This is indicated by the arrow in Figure \ref{fig7a}. \color{black}

The shot-to-shot repeatability in the near-orifice region \color{black} is shown in Figure \ref{fig7b}. High values (red) indicate large variations from one spray to the next, with low values (blue) indicating little variation. Both HFA152a and HFO1234ze(E) show greater shot-to-shot variation than HFA134a. The effect does not diminish with increasing ethanol content, but instead migrates from the core to the edges of the plume (indicated by arrows). Large shot-to-shot repeatability indicates variation in plume width and targeting that will increase total mouthpiece deposition. This may partly explain deposition changes in the APSD data (Figure \ref{fig2}, Table \ref{table:impactor}).


\begin{figure} \centering
\begin{subfigure}[b]{.49\textwidth} \centering
	\includegraphics[width=\textwidth,clip=true,trim=0.8cm 7.5cm 1.7cm 7.7cm]{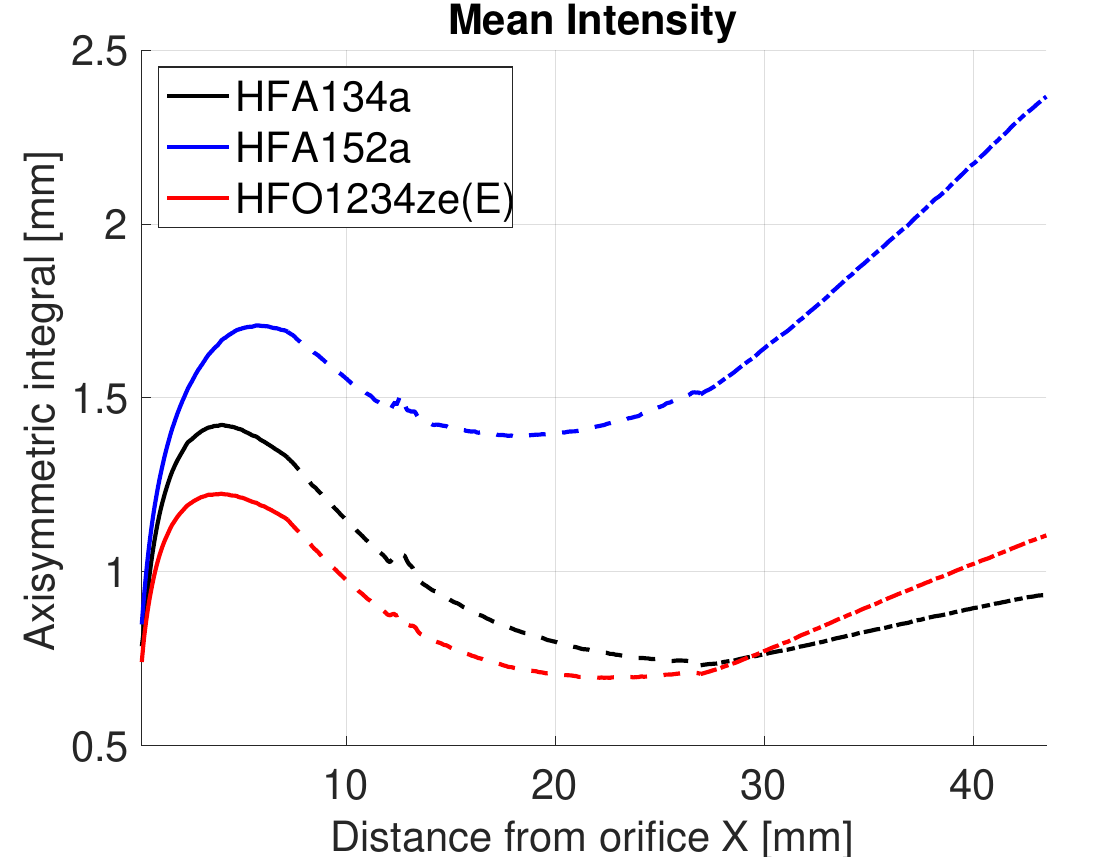}
	\caption{\label{fig8a}Combined radially integrated volumetric extinction $\left<\overline{I}_V\right>$.}
\end{subfigure}
\begin{subfigure}[b]{.49\textwidth} \centering
	\includegraphics[width=\textwidth,clip=true,trim=0.8cm 7.5cm 1.7cm 7.7cm]{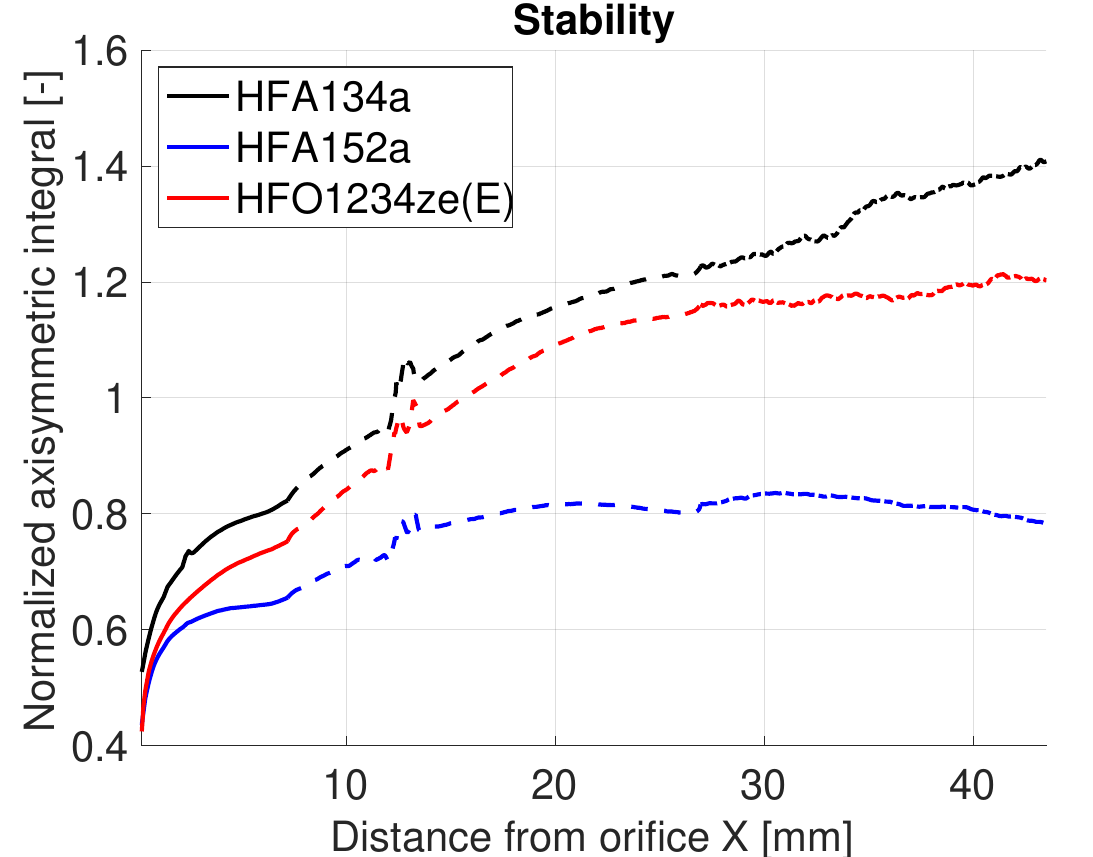}
	\caption{\label{fig8b}Ratio of radially integrated temporal stability to volumetric extinction $\Psi_{stab}/\left<\overline{I}_V\right>$.}
\end{subfigure}
\begin{subfigure}[b]{.49\textwidth} \centering
	\includegraphics[width=\textwidth,clip=true,trim=0.8cm 7.5cm 1.7cm 7.7cm]{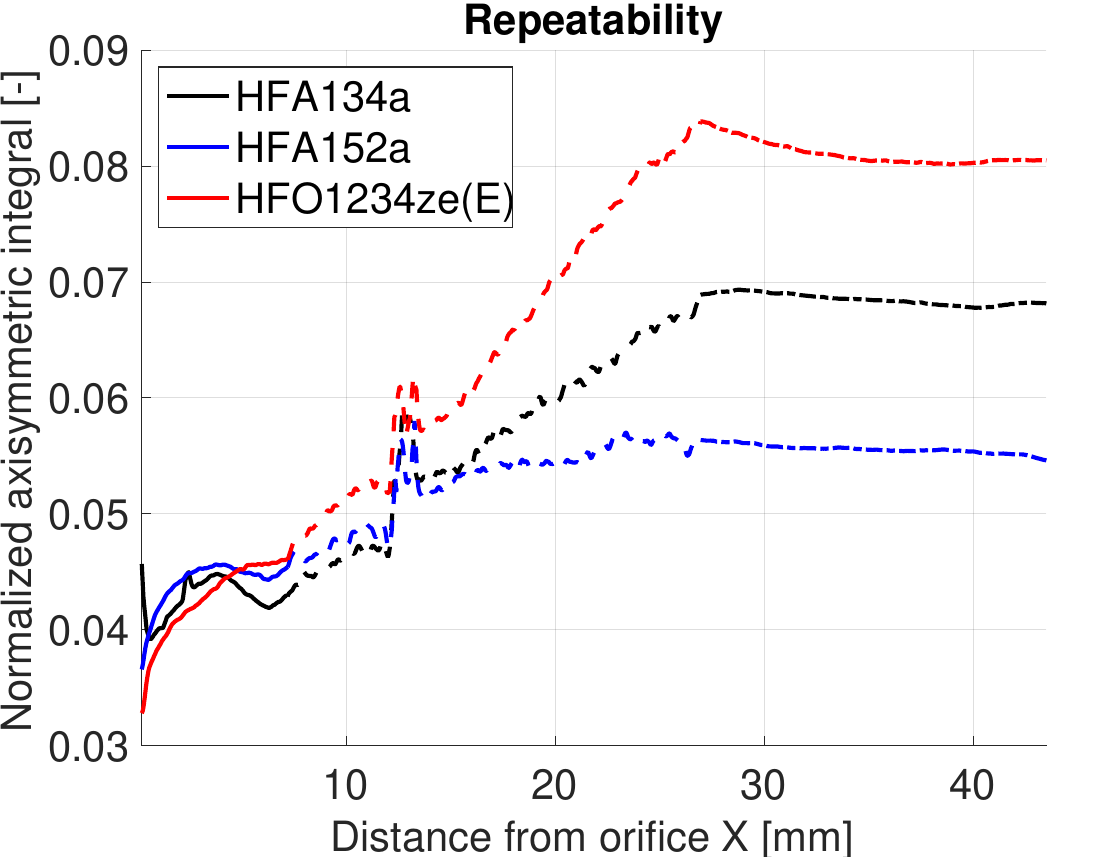}
	\caption{\label{fig8c}Ratio of radially integrated shot-to-shot repeatability to volumetric extinction $\Psi_{rep}/\left<\overline{I}_V\right>$.}
\end{subfigure}
\caption{\label{fig8}Radially integrated volumetric profiles for propellant-only placebo formulations.}
\end{figure}

The effect of stability and repeatability on spray performance can be quantified by radially integrating the profiles in Figure \ref{fig7} using similar logic to Equation \ref{eq:vol}, to obtain an axisymmetric evolution profile about the spray centerline $r_0$;
\begin{eqnarray}
\Psi_{stab}(x) = \int_{r_0(x)-r(x)}^{r_0(x)+r(x)} \left< \overline{I}_{stab} \right> (x,r) \, \mathrm{d} r \label{eq:psistab}\\
\Psi_{rep}(x) = \int_{r_0(x)-r(x)}^{r_0(x)+r(x)} \left< \overline{I}_{rep} \right> (x,r) \, \mathrm{d} r \label{eq:psirep}
\end{eqnarray}
The magnitude of $\Psi$ is arbitrary but its ratio with the volumetric extinction ($\left<\overline{I}_V\right>$, eqn. \ref{eq:vol}) provides useful insight into the underlying plume dynamics. If the mean spreading of the plume causes an equivalent stretching and scaling of the stability and repeatability statistics, and differences between propellants can be simply explained by this mean rescaling, we expect the ratio $\Psi/\left<\overline{I}_V\right>$ to be constant. Any rescaling would apply equally to both $\Psi$ and $\left<\overline{I}_V\right>$, and would thus cancel out when taken in ratio.

Figure \ref{fig8a} shows the volumetric extinction $\left<\overline{I}_V\right>$ for propellant-only placebo formulations. In order to obtain sufficient streamwise extent, three regions of interest (orifice, mouthpiece and ex-mouthpiece) have been joined together. These are indicated by the dashed lines. The first peak at $x \approx 5$ mm is caused by the initial rapid expansion driven by flash evaporation \cite{2012.Zeng}. The local minima at $x \approx 20$ mm is caused by constraining of the flow due to the presence of the mouthpiece, which is overcome by mixing further downstream as the profile increases again. All propellants show similar profiles with HFA152a again having a wider profile and HFO1234ze(E) narrower.

The normalized stability ratio $\Psi/\left<\overline{I}_V\right>$ for the temporal stability statistics are shown in Figure \ref{fig8b}. We note that this is not constant as expected but diverges immediately after the orifice (Fig. \ref{fig8b}). Both HFA152a and HFO1234ze(E) show reduced values but similar trends to HFA134a. This indicates that the main driver of differences in temporal stability is the orifice initial condition (solid lines).

The normalized shot-to-shot repeatability ratio $\Psi_{rep}/\left<\overline{I}_V\right>$ (Figure \ref{fig8c}) follows a similar profile near the orifice and diverges further downstream (Fig. \ref{fig8c}), becoming constant outside the mouthpiece. It is only in this region that the ratio becomes constant as expected. The abrupt changes in gradient are explained by the only factor which varies between the different measurements indicated by the dashed and solid lines; the cutting of the mouthpiece near the orifice to obtain optical access. This indicates that shot-to-shot repeatability is strongly determined by mouthpiece shape and size. \color{black}

\begin{figure} \centering
\begin{subfigure}[b]{.49\textwidth} \centering
	\includegraphics[width=\textwidth,clip=true,trim=0.8cm 7.5cm 1.7cm 7.7cm]{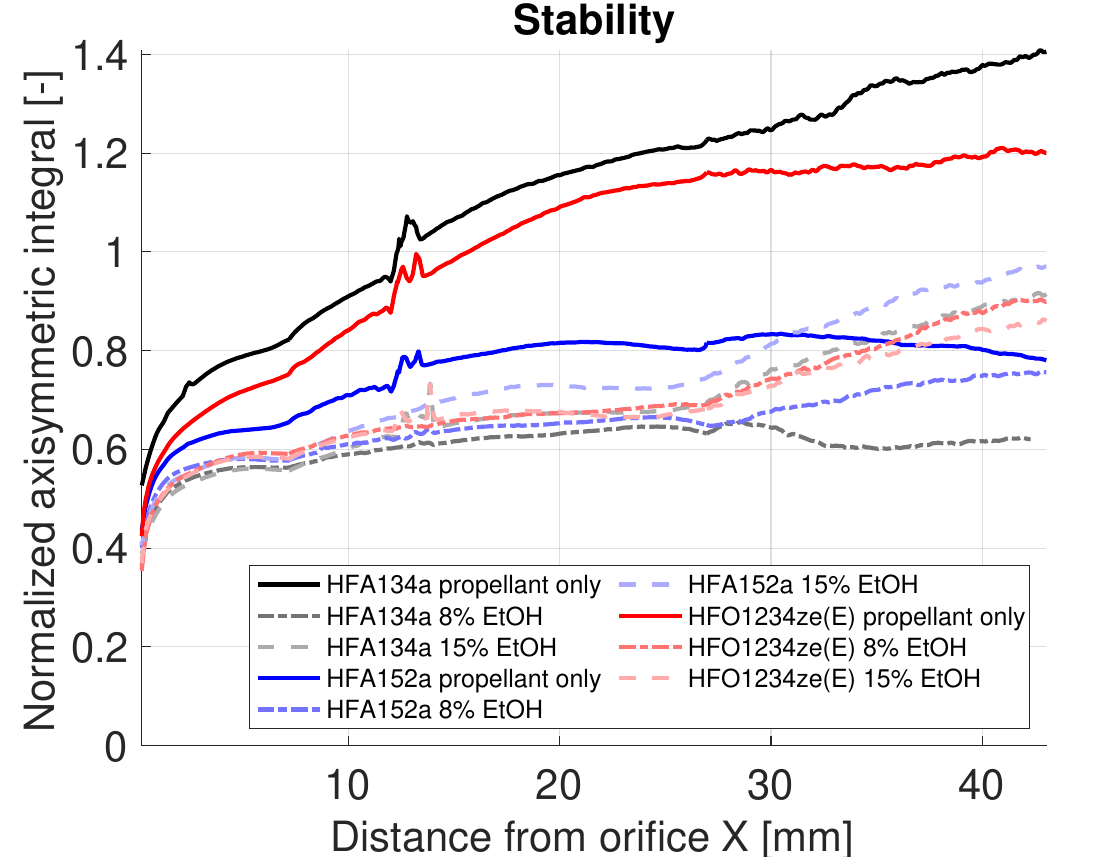}
	\caption{\label{fig9a}Ratio of radially integrated temporal stability to volumetric extinction $\Psi_{stab}/\left<\overline{I}_V\right>$.}
\end{subfigure}
\begin{subfigure}[b]{.49\textwidth} \centering
	\includegraphics[width=\textwidth,clip=true,trim=0.8cm 7.5cm 1.7cm 7.7cm]{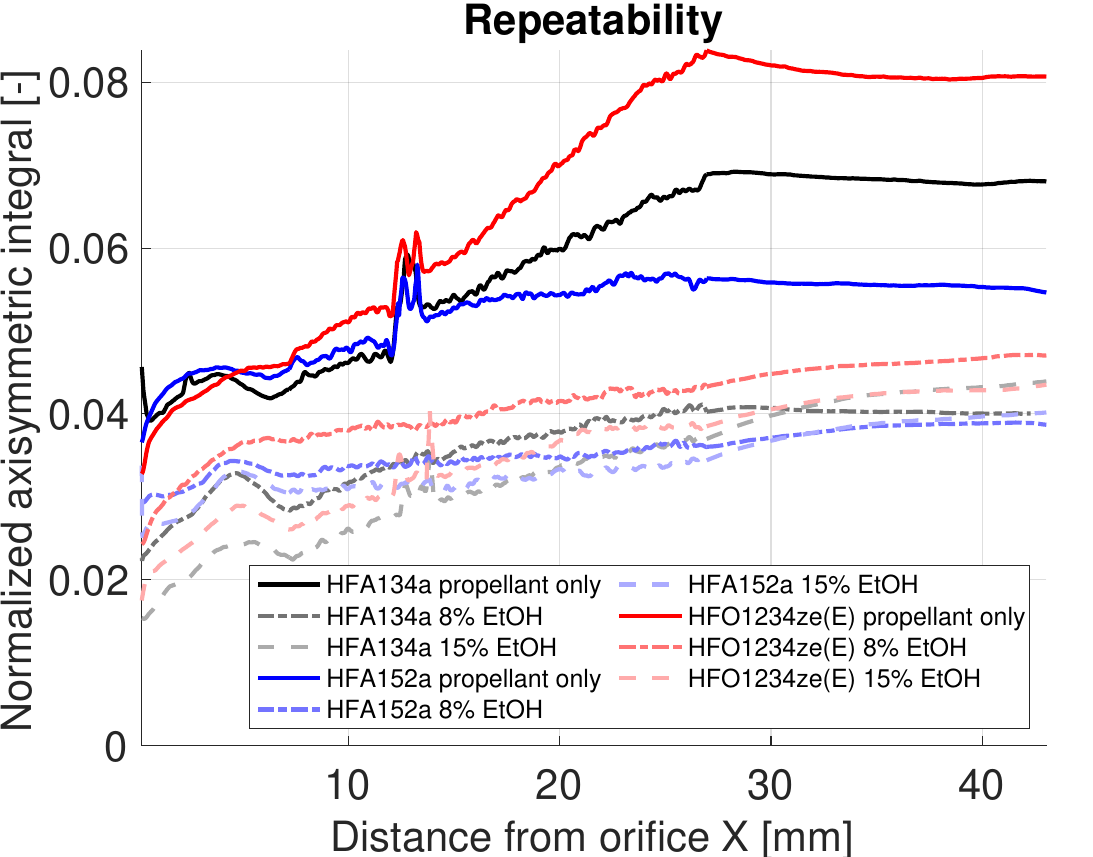}
	\caption{\label{fig9b}Ratio of radially integrated shot-to-shot repeatability to volumetric extinction $\Psi_{rep}/\left<\overline{I}_V\right>$.}
\end{subfigure}
\caption{\label{fig9}Radially integrated volumetric profiles for propellant-only placebos, 8\% and 15\% w/w ethanol solution formulations of 2.0 mg/mL BDP.}
\end{figure}

The differences between formulations are expected to be strongest in the cosolvent-free formulations shown in Figure \ref{fig8}. Figure \ref{fig9} shows the effect of adding ethanol with the propellant-only placebo compared to 8\% and 15\% w/w ethanol solutions of 2.0 mg/mL BDP (indicated by dashed lines). As expected, the profiles flatten and differences are far less marked with increasing ethanol content. As ethanol is added, the spray trends towards the ideal scenario described above where the stability and repeatability are simply described by stretching and scaling of the mean extinction. This is likely due to the greatly reduced volatility of ethanol; flashing, convective mixing and evaporative effects are reduced.

\section{Conclusions}

A combination of traditional aerodynamic particle size distribution measurements, droplet size distribution data and novel high-speed imaging measurements have been used to elucidate the differences between solution formulations containing HFA134a, HFA152a and HFO1234ze(E) propellants with matched hardware across various cosolvent levels and drug concentrations. The primary advantage of the high-speed imaging technique described here beyond conventional spray pattern and plume geometry \cite{Baxter.2022} is the ability to extract detailed temporal stability and shot-to-shot repeatability statistics over very large data sets with high spatial and temporal resolution. This reveals the regions and factors that drive differences between the formulations, and potential routes to tuning spray structure to achieve targeted outcomes.

HFA152a solution formulations showed reduced FPD relatively to HFA134a and a significant increase in actuator, coupler and throat deposition. HFO1234ze(E)-containing solutions showed an improved FPD with no significant change in deposition. In the absence of cosolvent, HFO1234ze(E) appears to exhibit a notable decrease in delivered dose which is mitigated with the addition of ethanol. Both HFA152a and HFO1234ze(E) have larger particle and droplet MMAD, with the increases being attributable to their chemicophysical property changes which give rise to larger initial droplet size at the orifice.

High-speed imaging revealed that HFA152a sprays are consistently wider and HFO1234ze(E) sprays are consistently narrower than HFA134a under all conditions. It was observed that near-orifice effects were correlated with vapour density due to flash evaporation. The lower density of HFA152a relative to HFA134a gives rise to greater expansion of the vapour phase and a wider, less stable spray and increased deposition. Ex-mouthpiece effects were more strongly correlated with formulation mixture density through turbulent mixing. In an isenthalpic flash scenario, HFA134a and HFO1234ze(E) have similar mixture densities and thus trend toward similar ex-mouthpiece spray profiles while HFA152a remains wider as its density is lower.

Both HFA152a and HFO1234z(E) formulations show increased transient behaviour (greater temporal instability) relative to HFA134a. The differences are moderated with the addition of ethanol, however HFA152a requires more ethanol to achieve the same degree of stabilisation.  Both low-GWP propellants also show increased shot-to-shot variability (repeatability). Addition of ethanol modifies the repeatability but does not eliminate the effect. 

Changes in temporal stability were found not to scale with the mean spray profile but rather depended on the initial condition near the orifice, which creates an initial divergence in plume dynamics that then propagates outward, with greater instability at the orifice naturally leading to greater instability further downstream. \color{black}This suggests that plume stability may be adjusted for new low-GWP formulations primarily through orifice geometry modification. Addition of ethanol is a secondary route, but is less effective in HFA152a formulations. Increasing ethanol content may also lead to other undesired changes in droplet and particle size, vapor pressure and solubility. 

Changes in.shot-to-shot repeatability between propellant were found to be significantly affected by the mouthpiece geometry, as cutting the mouthpiece to obtain optical access led to large and discontinuous changes in the normalized repeatability statistics. This suggests that repeatability and delivered dose uniformity challenges, which may not be easily addressed with cosolvent or formulation changes, might be better treated by adjustment of the mouthpiece geometry.

These early results suggest that a range of hardware options will need to be considered for various applications.  \color{black}From a plume structure and performance perspective the differences between propellants are sufficiently small as to be able to be managed through a combination of formulation, orifice and mouthpiece geometry changes. This holds promise for the use of in vitro bioequivelance data to support the rapid development of low-GWP replacement pMDIs using more detailed plume structure and geometry measurements as part of a weight-of-evidence based approach \cite{jeanneret}.

\section*{Acknowledgements}

The authors acknowledge the support of the Australian Research Council (grants LP190100938 \& DP200102016). We gratefully acknowledge the computing resources provided on \textit{Bebop}, a high-performance computing cluster operated by the Laboratory Computing Resource Center at Argonne National Laboratory. The authors also acknowledge the use of the National Computational Infrastructure (NCI), which is supported by the Australian Government.


\singlespacing

\bibliographystyle{unsrt}
\bibliography{imagingPaper}

\section*{Appendix A: Nomenclature}

\begin{tabular}{ll}
$c_p$ & Specific heat capacity \\
$D$ & Orifice exit diameter \\
$d$ & Droplet diameter \\
$d_{32}$ & Droplet Sauter Mean diameter (Eqn. \ref{eq:d32}) \\
$d^*$ & Theroetical minimum stable droplet size as determined by surface tension (Eqn. \ref{eq:WeMin})\\
GSD & Geometric standard deviation\\
$h_{fg}$ & Latent heat of vapourisation\\
$I$ & Normalized light extinction in spray \\
$I_V$ & Volume-integrated light extinction in spray (Eqn. \ref{eq:vol})\\
$I_{stab}$ & Temporal stability of the normalized light extinction (Eqn. \ref{eq:stab}) \\
$I_{rep}$ & Shot to shot repeatability of the normalized light extinction (Eqn. \ref{eq:rep}) \\
$\overline{I}$ & Time-average light extinction \\
$\left< I \right>$ & Ensemble-average light extinction\\
Ja & Jakob number; ratio of sensible to latent heat (Eqn. \ref{eq:Ja})\\
MMAD & Mass mean aerodynamic diameter \\
$n$ & Number of actuations\\
$N$ & Number of repeated trials \\
$P$ & Pressure\\
$P_{sat}$ & Saturated vapor pressure\\
$P_{amb}$ & Ambient (environmental) pressure\\
$r$ & radial distance from orifice \\
$r_0$ & spray centerline radius \\
Re & Reynolds number ; ratio of inertial to viscous forces  \\
$S$ & Droplet surface area\\
$s_i, s_t$ & Sample standard deviation in space and time\\
$T$ & Temperature \\
$T_0$ & Ambient temperature \\
$T_{sat}$ & Saturation temperature of the formulation at atmospheric pressure \\
$t$ & time\\
$t_0, t_1$ & start and end time of the steady-state period of the spray\\
$\overline{U}$ & Average (bulk) orifice velocity, flow rate divided by orifice cross-section area \\
$V$ & Droplet volume\\
We & Weber number ; ratio of intertial to surface tension forces \\
$x$ & axial distance from orifice\\
$\rho$ & Density\\
$\rho_{mix}$ & Formulation mixture density at the orifice \\
$\mu$ & Dynamic viscosity \\
$\sigma$ & Surface tension \\
$\Psi$ & Axisymetrically-integrated stability (Eqn. \ref{eq:psistab}--\ref{eq:psirep})
\end{tabular}

\clearpage
\section*{Supplementary Material}

\begin{figure}[h]
\includegraphics[width=\textwidth,height=8cm,clip=true,trim=1cm 7mm 7mm 1cm]{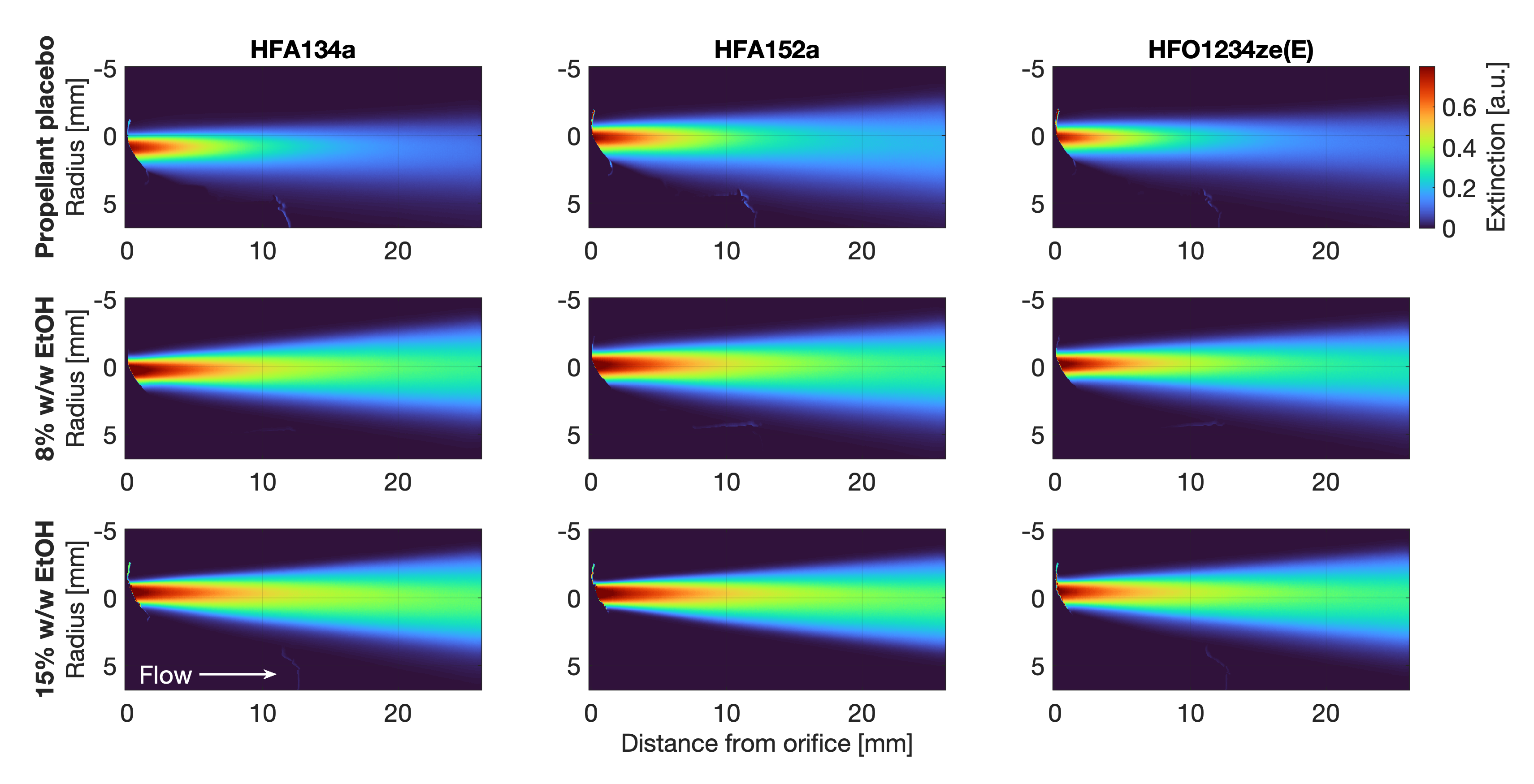}
\caption{\label{figS1}Spatial structure of plumes for all formulations  in the near-orifice region \color{black} for time-average, ensemble-average mean extinction (Eqn. \ref{eq:mean}). Flow is left-to-right. The horizontal axes are distance from orifice ($x$) and vertical axes are radius ($r$), in mm.}
\end{figure}

Figure \ref{figS1} shows the time-average, ensemble-average light extinction for the propellant formulations considered in this study (Equation \ref{eq:mean}). The mean profiles are unremarkable and show relatively little difference. The majority of the differences between propellants are found in the integrated light extinction and higher-order statistics such as temporal stability and shot-to-shot repeatability.

\end{document}